%% file: main.tex
\documentclass[10pt,journal]{IEEEtran}

\ifCLASSOPTIONcompsoc

\else
\fi
\ifCLASSINFOpdf
 \else
\fi

\usepackage{graphicx, url}
\usepackage[noadjust]{cite}
\usepackage{balance}
\usepackage{subfigure,captcont,wrapfig, setspace}
\usepackage{comment, ragged2e}
\usepackage{subfigure, color, colortbl}
\usepackage{amssymb,amsmath}
\usepackage[linesnumbered, ruled, vlined]{algorithm2e}
\definecolor{blue}{rgb}{0.0,0.0,1}

\definecolor{red}{rgb}{1,0.0,0.0}

\newtheorem{theorem}{Theorem}
\newtheorem{lemma}{Lemma}

\begin{document}
%
\title{Integrating Low-Power Wide-Area Networks for Enhanced Scalability and Extended Coverage}
%
%
%
%

\author{Mahbubur~Rahman and
        Abusayeed~Saifullah
\IEEEcompsocitemizethanks{\IEEEcompsocthanksitem Mahbubur Rahman and Abusayeed Saifullah are with the Department
of Computer Science, Wayne State University, Detroit,
MI, 48202. \protect\\
E-mail: r.mahbub@wayne.edu, saifullah@wayne.edu
}}

\IEEEtitleabstractindextext{%
\justify{
\begin{abstract}

Low-Power Wide-Area Networks (LPWANs) are evolving as an enabling technology for Internet-of-Things (IoT) due to their capability of communicating over long distances at very low transmission power.
Existing LPWAN technologies, however, face limitations in meeting scalability and covering very wide areas which make their adoption challenging for future IoT applications, especially in infrastructure-limited rural areas.
To address this limitation, in this paper, we consider achieving scalability and extended coverage by integrating multiple LPWANs. 
{\bf\slshape SNOW (Sensor Network Over White Spaces)}, a recently proposed LPWAN architecture over the TV white spaces, has demonstrated its advantages over  existing LPWANs in performance and  energy-efficiency.  
In this paper, we propose to scale up LPWANs through a seamless  integration of multiple SNOWs which enables concurrent inter-SNOW and intra-SNOW communications.  
We then formulate the tradeoff between scalability and inter-SNOW interference as a constrained optimization problem whose objective is to  maximize scalability by managing white space spectrum sharing across multiple SNOWs. We also prove the NP-hardness of this problem. 
To this extent, We propose an intuitive polynomial-time heuristic algorithm for solving the scalability optimization problem which is highly efficient in practice. For the sake of theoretical bound, we also propose a simple polynomial-time $\frac{1}{2}$-approximation algorithm for the scalability optimization problem. 
Hardware experiments through deployment in an area of (25x15)km$^2$ as well as large scale simulations demonstrate the effectiveness of our algorithms and feasibility of achieving scalability through seamless integration of SNOWs with high reliability, low latency, and energy efficiency.
\end{abstract}}

}

\maketitle

\IEEEdisplaynontitleabstractindextext
\IEEEpeerreviewmaketitle

\input{introduction}

\input{related}

\input{snow_overview}

\input{system_model}

\input{inter_snow_comm}

\input{technical}

\input{experiment}

\input{simulation}

\section{Conclusions}\label{sec:conclusions}

LPWANs represent a key enabling  technology for IoT that offer long communication range at low power. While many competing LPWAN technologies have been developed recently, they still face limitations in meeting scalability and covering much wider area. Such limitations make the adoption of LPWANs challenging for future IoT applications, especially in  infrastructure-limited rural areas. In this paper, we have addressed this challenge by integrating multiple LPWANs for enhanced scalability and extended coverage.
Specifically, we have proposed to scale up LPWANs through a seamless  integration of multiple SNOWs that enables concurrent inter-SNOW and intra-SNOW communications.  We have then formulated the tradeoff between scalability and inter-SNOW interference as a scalability optimization problem, and have proved its NP-hardness. Consequently, we have proposed a polynomial-time greedy heuristic that is highly effective in experiments as well as a polynomial-time $1/2$-approximation algorithm. Testbed experiments as well as large scale simulations demonstrate the feasibility of achieving scalability through our proposed integration of SNOWs with high reliability, low latency, and energy efficiency.

\section*{Acknowledgment}
This work was supported by NSF through grants CNS-1742985 and CAREER-1846126.

\bibliographystyle{IEEEtran}
\bibliography{IEEEabrv,whitespacebib}

\begin{IEEEbiography}[{\includegraphics[width=25mm,height=32mm,clip,keepaspectratio]{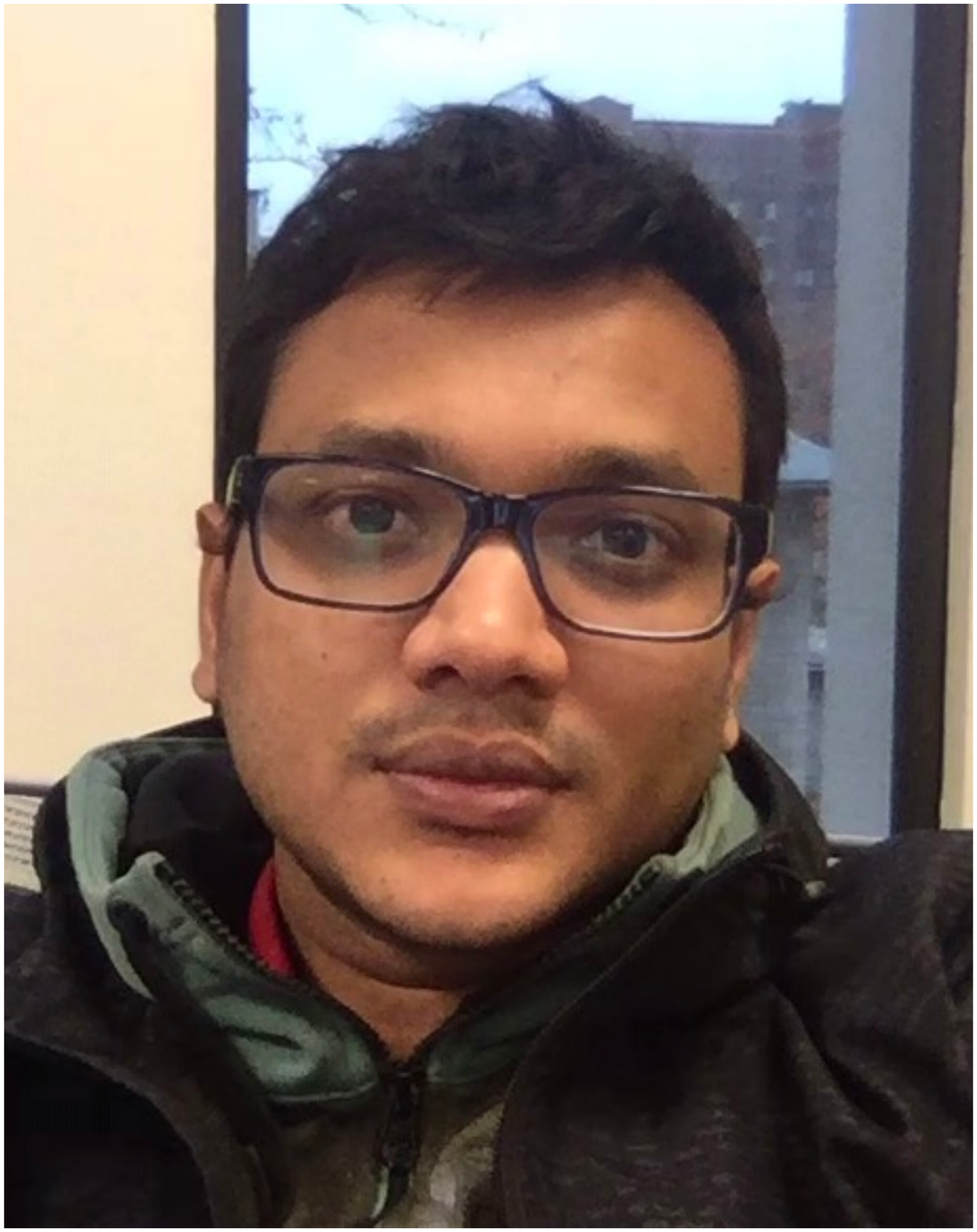}}]{Mahbubur Rahman}
received the bachelor’s degree in computer science and engineering from Bangladesh University of Engineering and Technology, Dhaka, Bangladesh, in 2012. He is currently pursuing the Ph.D. degree with the Department of Computer Science, Wayne State University. He has co-primary-authored a paper that was nominated for the best paper at ACM SenSys 2016. His research interests include low-power wide-area networks, Internet of Things, and cyber-physical systems.
\end{IEEEbiography}

\vspace{-0.5in}
\begin{IEEEbiography}[{\includegraphics[width=25mm,height=32mm,clip,keepaspectratio]{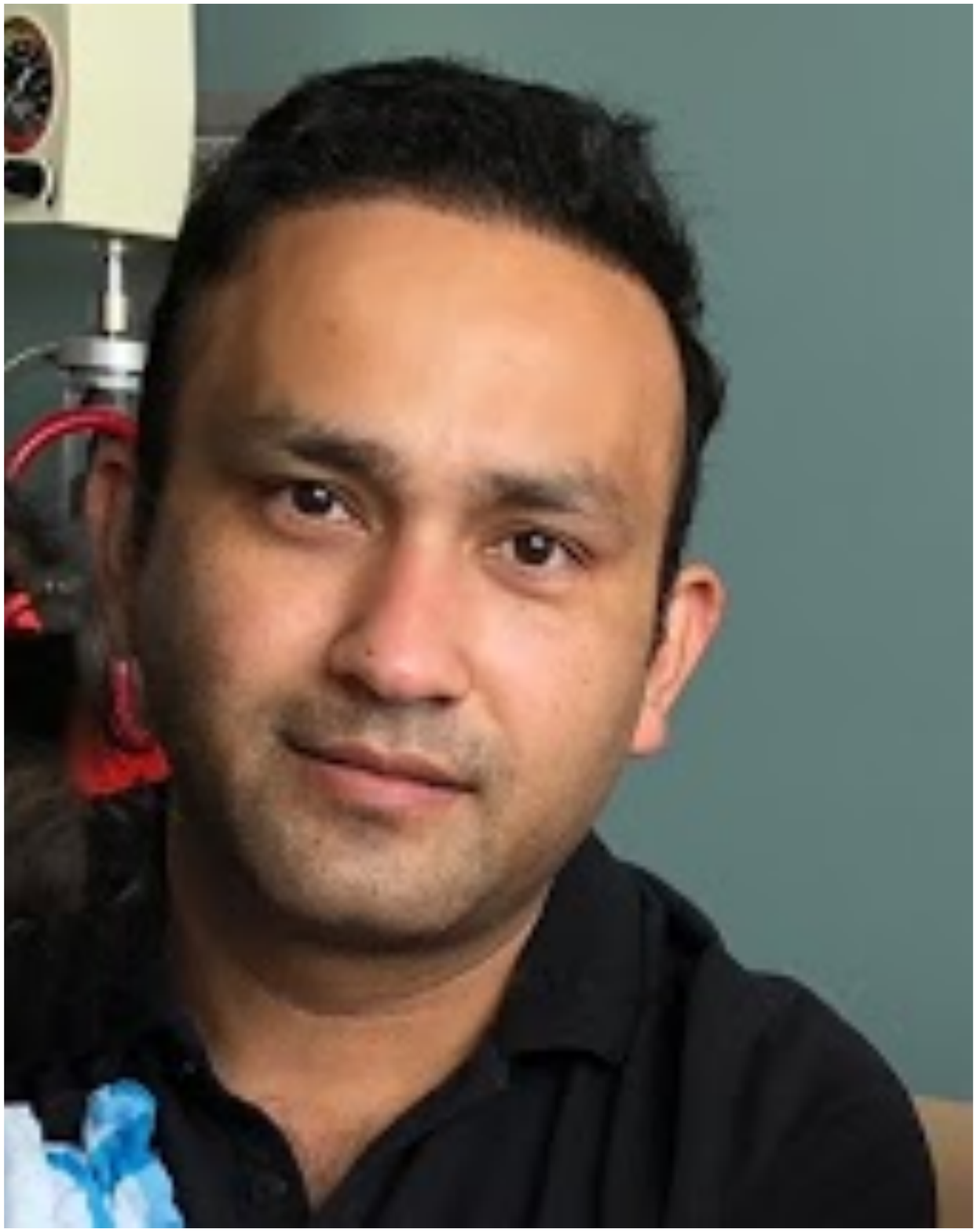}}]{Abusayeed Saifullah}
is an assistant professor of the Computer Science Department at Wayne State University. He received PhD in Computer Science and Engineering with Turner Dissertation Award from Washington University in St Louis in 2014. His research primarily concerns Internet-of-Things, cyber-physical systems, real-time systems, embedded systems, and low-power wide-area networks. He received 7 Best Paper Awards/Nominations in highly competitive conferences including ACM SenSys (2016 nomination), IEEE RTSS (2019, 2014, 2011), IEEE ICII (2018), and IEEE RTAS (2012 nomination). He also received multiple young investigator awards including the CAREER award (2019) and the CRII award (2016) of the National Science Foundation (NSF). He is serving as the program chair of IEEE ICESS 2020, served as a track chair of IEEE ICCCN 2019 and as a program committee member for various conferences including ACM SenSys, IEEE RTSS, ACM/IEEE IoTDI, IEEE RTAS, ACM/IEEE ICCPS, ACM MobiHoc, IEEE  INFOCOM, EWSN, and ACM IWQoS. He also served as a guest editor of IEEE Transactions on Industrial Informatics, and is currently an editor of Elsevier Pervasive and Mobile Computing journal. 
\end{IEEEbiography}

\end{document}

%% file: introduction.tex
\section{Introduction}\label{sec:introduction}
To overcome the range limit and scalability challenges in traditional wireless sensor networks (WSNs), Low-Power Wide-Area Networks (LPWANs) are emerging as an enabling  technology for Internet-of-Things (IoT).   Due to their escalating demand, LPWANs are gaining momentum, with multiple competing technologies being developed including  LoRaWAN, SigFox, IQRF, RPMA (Ingenu), DASH7, 
Weightless-N/P in the ISM band; and 
EC-GSM-IoT, NB-IoT,  LTE Cat M1  (LTE-Advanced Pro), and 5G in the licensed cellular band (see survey~\cite{ismail2018low}). 
In parallel, to avoid the crowd of the limited ISM band and the cost of the licensed band, we developed {\bf\slshape  SNOW (Sensor Network Over White Spaces)}, an LPWAN architecture  to support wide-area WSN by exploiting the TV white spaces~\cite{snow_ton, snow, snow2}. 
{\slshape White spaces} refer to the allocated but locally unused TV channels, and can be used by unlicensed devices as secondary users. Unlicensed devices need to either sense the medium or consult with a cloud-hosted geo-location database before transmitting~\cite{whitespaceSurvey}.
Thanks to their lower frequencies (54--862MHz in the US), white spaces have excellent propagation characteristics over long distance and obstacles.
While their potentials have been explored mostly for broadband access (see survey~\cite{whitespaceSurvey}), our 
design and experimentation demonstrated the potential of SNOW to enable asynchronous, low power, bidirectional, and massively concurrent communications between numerous sensors and a base station (BS) directly over long distances~\cite{snow, snow2, snow_ton}. 

Despite their promise, existing LPWANs face challenge in very large-area (e.g., city-wide) deployment~\cite{charm, choir}. Without line of sight, communication range of LoRaWAN, a leading LPWAN technology that is commercially available, is  short, especially in indoors ($<$100m while its specified urban range is 2--5km)~\cite{marcelis2017dare}.  Its performance  drops sharply as the number of nodes grows, supporting only 120 nodes per 3.8 hectares~\cite{bor2016lora} which is not sufficient to meet the future IoT demand.
Apart from these scenarios, applications like 
agricultural IoT,  oil-field monitoring,  smart and connected rural communities would require much wider area coverage~\cite{ismail2018low, whitespaceSurvey}.  In this paper, we address this {\bf challenge} and propose LPWAN scalability by integrating multiple LPWANs.

Most LPWANs are limited to star topology, and rely on wired infrastructure (e.g., cellular LPWANs) or Internet (e.g., LoRaWAN) to integrate multiple networks to cover large areas. Lack of infrastructure (also raised in a hearing before the US Senate~\cite{monsanto}) hinders their adoption to enable rural and remote area applications such as {\bf agricultural IoT}  and {\bf industrial IoT}  (e.g., for oil/gas field) that may cover hundreds of square kms. According to the Department of Agriculture, $<20\%$ farmers can afford the cost of manual sensor data collection for smart farming \cite{doa}. 
Industries like Microsoft~\cite{vasisht2017farmbeats},  Monsanto~\cite{monsanto},  and many \cite{ismail2018low, whitespaceSurvey}  are now promoting agricultural IoT.  Monitoring a large oil-field (e.g., 74x8km$^2$ East Texas Oil-field~\cite{texasof})  needs to connect tens of thousands of sensors ~\cite{whitespaceSurvey}.  Such agricultural IoT and industrial IoT can be enabled by integrating multiple LPWANs specially SNOWs due to abundant white spaces.
Similar integration may also be needed in a smart city deployment for extended coverage or for running different applications on different LPWANs.

In this paper, we address the above scalability challenge  by integrating multiple  SNOWs that are under the same management/control. Such an integration raises several concerns. First, we have to design a  protocol to enable inter-SNOW communication, specially peer-to-peer communication (when a node in one SNOW wants to communicate with a node in a different SNOW). Second, since multiple coexisting SNOWs can interfere each other, thus affecting the scalability,   it is critical to handle the tradeoffs between scalability and inter-SNOW interference.  Specifically, we make the following  novel contributions.

\begin{itemize}
\item We propose to scale up LPWAN through seamless  integration of multiple SNOWs that enables concurrent inter- and intra-SNOW communications.  This is done by exploiting the characteristics of the SNOW physical layer.

\item We then formulate the tradeoff between scalability and inter-SNOW interference as a constrained optimization problem whose objective is to  maximize scalability by managing white space spectrum sharing across multiple SNOWs, and prove its NP-hardness.

\item We propose an intuitive polynomial-time heuristic for solving the scalability optimization problem which is highly efficient in practice.

\item For the sake of analytical performance bound,  we also propose a simple polynomial-time approximation algorithm with an approximation ratio of $\frac{1}{2}$.

\item We implement the proposed SNOW technologies in GNU Radio~\cite{gnuradio} using Universal Software Radio Peripheral (USRP) devices~\cite{usrp}. We perform experiments by deploying 9 USRP devices in an area of (25x15)km$^2$ in Detroit, Michigan.  We also perform large scale simulations in NS-3~\cite{ns3}. Both experiments and simulations demonstrate the feasibility of achieving scalability through seamless integration of SNOWs allowing  concurrent intra- and inter-SNOW communications with high reliability, low latency, and energy efficiency while using our heuristic and approximation algorithms. Also, simulations show that SNOW cluster network can connect thousands of sensors over tens of kilometers of geographic area.
\end{itemize}


In the rest of the paper, Section~\ref{sec:related} presents related work. Section~\ref{sec:snow_overview} gives an overview of SNOW. Section~\ref{sec:system_model} explains the system model.  
Section~\ref{sec:p2p} describes our inter-SNOW communication technique. Section~\ref{sec:technical} formulates the scalability optimization problem for integration, proves its NP-hardness, and presents the heuristic and the approximation algorithm. Section~\ref{sec:implementation} explains the implementation of our network model. Section~\ref{sec:eval} presents our experimental and simulation results. Finally, Section~\ref{sec:conclusions} concludes the paper.

%% file: related.tex
\section{Related Work}\label{sec:related}

The LPWAN technologies are still in their infancy with some still being developed (e.g., 5G, NB-IoT, LTE Cat M1, Weightless-P), some having only uplink capability (e.g., SigFox, Weightless-N), while, for some, there is still no publicly available documentation (e.g., SigFox)~\cite{ismail2018low, whitespaceSurvey}. Thus, developing generalized techniques to address integration is not our focus. Instead, we propose an integration of multiple SNOWs in the white spaces for scaling up, the insights of which may also be extended to other LPWANs in the future.
To cover a wide area, LoRaWAN integrates multiple gateways through the Internet~\cite{lorawan}. Cellular networks do the same relying on wired infrastructure~\cite{channel_cellular}.  Rural and remote areas lack such infrastructure.  Wireless integration that we have considered in this paper can be a solution for both urban and rural areas.

While our integration may look similar to channel allocation in traditional tiered/clustered and centralized/distributed multi-channel networks~\cite{si2010overview, ko2007distributed, raniwala2005architecture, kodialam2005characterizing, marina2010topology, 80211mesh, rad2006joint, gurewitz2007cooperative, aryafar2008distance, li2019bio, zhao2018simple}, it is a {\bf\slshape conceptually different} problem with new challenges.
{\bf First,} in traditional networks, the links operate on predefined fixed-bandwidth channels. In contrast, in integrating multiple SNOW networks we have to find proper bandwidths for all links and they are inter-dependent and can be different.  
{\bf Second,} SNOW integration involves assigning a large number of subcarriers to each BS allowing some degree
of overlaps among interfering BSs for enhanced scalability.
{\bf Finally,} through integration, we have to retain massive parallel communication (between a SNOW BS and its numerous nodes) and concurrent inter- and intra-SNOW communications~\cite{snow_integration, rahman2018demo}.
Hence, traditional channel allocation for wireless networks~\cite{channel_wireless}, WSN~\cite{channel_wsn, leach}, or cognitive radio networks~\cite{cognitive_channel_survey} cannot be used in SNOW integration.
In regard to the white space networking, the closest work to ours is~\cite{gameapproach} which considers multiple WiFi-like 
networks in white spaces, where all users have access to white
space database, and every access point (AP) chooses a single
channel, the problem thus is different from our integration.

%% file: snow_overview.tex
\section{An Overview of SNOW}\label{sec:snow_overview}
\begin{figure}[!htb]
\centering
\includegraphics[width=0.45\textwidth]{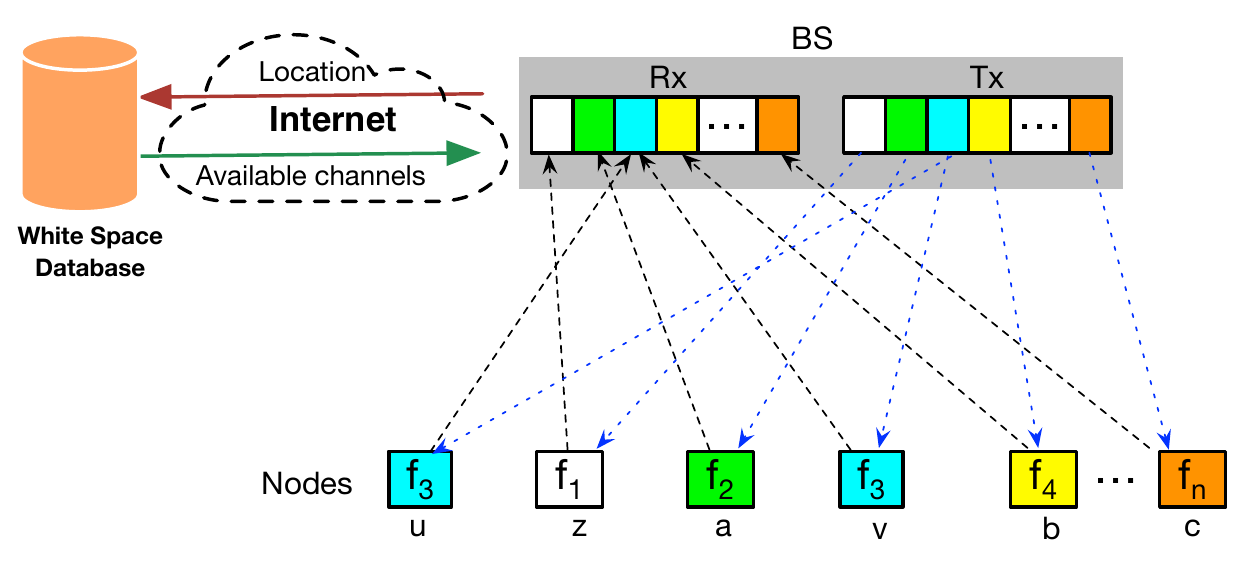}
\caption{SNOW architecture with dual radio BS and subcarriers.}
\label{fig:dualradio}
\end{figure}
Here we provide a brief overview of the design and architecture of a {\bf single} SNOW that we developed in~\cite{snow, snow2, snow_ton}.
SNOW is an asynchronous,  long range, low power WSN platform to operate over TV white spaces. A SNOW node has a single half-duplex narrowband radio.  
Due to long transmission (Tx) range, the nodes are directly connected to the BS and vice versa (Figure~\ref{fig:dualradio}). SNOW thus forms a star topology.    
The BS determines white spaces in the area by accessing a cloud-hosted database  through the Internet. 
Hence, it does not check on the incumbents or evaluate cross-technology interference.
The nodes  are power constrained and not directly connected to the Internet.  They do not do spectrum sensing or cloud access.
The BS uses a wide channel split into orthogonal subcarriers.   As shown in Figure~\ref{fig:dualradio}, the BS uses two radios, both operating on the same spectrum --  one for only transmission (called {\bf\slshape Tx radio}),  and the other for only reception (called {\bf\slshape Rx radio}). Such a dual-radio of the BS allows concurrent
bidirectional communications in SNOW. 
We implemented SNOW on USRP (universal software radio peripheral) devices~\cite{usrp} using GNU Radio~\cite{gnuradio}.  
The implementation has been made {\bf open-source}~\cite{snow_bs, snow_cots}. A short video demonstrating how SNOW works is also available in YouTube~\cite{snow_demo, ismail2018demo}. In the following, 
we provide a brief overview of the SNOW physical layer
(PHY) and the Media Access Control (MAC) layer. A full description of this design is available in~\cite{snow_ton}.

\subsection{SNOW PHY Layer}\label{sec:phy}

A key design goal of SNOW is to achieve high scalability by exploiting wide spectrum of white spaces. Hence, its PHY is designed based on a 
{\bf D}istributed implementation of {\bf OFDM} for multi-user access, called {\bf D-OFDM}. D-OFDM  splits a wide spectrum into numerous narrowband orthogonal subcarriers enabling parallel data streams to/from numerous distributed nodes from/to the BS.  A subcarrier bandwidth is in kHz (e.g., 50kHz, 100kHz, 200kHz, or so depending on packet size and needed bit rate). Narrower bands have lower bit rate but longer range, and consume less power~\cite{snow}. The nodes transmit/receive on orthogonal subcarriers, each using one. A subcarrier is modulated using Binary Phase Shift Keying (BPSK) or Amplitude Shift Keying (ASK).  If the BS spectrum is split into  $n$  subcarriers,  it can receive from $n$ nodes simultaneously using a single antenna. Similarly, it can transmit different data on different subcarriers through a single transmission. The BS can also use fragmented spectrum. This design is different from MIMO radio adopted in various wireless domains including IEEE 802.11n~\cite{mimo} as they rely on multiple antennas to enable the same. 


While OFDM has been adopted  for multi-access in the forms of OFDMA  and SC-FDMA in various  broadband (e.g., WiMAX~\cite{wimax}) and cellular  (e.g., LTE) technologies~\cite{3gpp, scfdma, OFDMAWiMAX}, they rely on strong time synchronization which is very costly for low-power nodes. We adopted OFDM for the first time in WSN design and without requiring time synchronization. D-OFDM enables multiple packet receptions that are transmitted asynchronously from different nodes which was possible as WSN needs low data rate and short packets. Time synchronization is avoided by extending the symbol duration (repeating a symbol multiple times) and sacrificing bit rate. The effect is similar to extending cyclic prefix (CP) beyond what is required to control  inter-symbol interference (ISI). CPs of adequate lengths have the effect of rendering asynchronous signals to appear orthogonal at the receiver, increasing guard-interval. As it reduces data rate, D-OFDM is suitable for LPWAN. Carrier frequency offset (CFO) is estimated using training symbols when a node joins the network on a subcarrier (right most) whose overlapping subcarriers are not used. Using this CFO, it is  determined on its assigned subcarrier and compensated for using traditional method to mitigate ICI.

\subsection{SNOW MAC Layer}

The BS spectrum  is split into  $n$ overlapping orthogonal subcarriers -- $f_1, f_2, \cdots, f_n$ -- each of equal width. Each node is assigned one subcarrier. When the number of nodes is no greater than the number of subcarriers, every node is assigned a unique subcarrier.  Otherwise, a subcarrier is shared by more than one node. 
The nodes that share the same subcarrier will contend for and access it using a CSMA/CA (Carrier Sense Multiple Access with Collision Avoidance) policy. The subcarrier assignment by the BS minimizes the interference and contention between the nodes. As long as there is an option, the BS thus tries to assign different subcarriers to the nodes that are hidden to each other.

The subcarrier allocation is done by the BS.
The nodes in SNOW use a lightweight CSMA/CA protocol for transmission that uses a static interval for random back-off like the one used in TinyOS~\cite{tinyos} . 
Specifically, when a node has data to send, it wakes up by turning its radio on. Then it performs a random back-off in a fixed {\slshape initial back-off window}.  When the back-off timer expires, it runs CCA (Clear Channel Assessment) and if the subcarrier is clear, it transmits the data. If the subcarrier is occupied, then the node makes a random back-off in a fixed {\slshape congestion back-off window}. After this back-off expires, if the subcarrier is clean the node transmits immediately. This process is repeated until it makes the transmission. The node then can go to sleep again.

The nodes can autonomously transmit, remain in receive (Rx) mode, or sleep. 
Since D-OFDM allows handling asynchronous Tx and 
Rx, the link layer can send acknowledgment (ACK) for any transmission in either direction.   As shown in Figure~\ref{fig:dualradio}, both radios of the BS use the same spectrum and subcarriers - the subcarriers in the Rx radio are for receiving while those in the Tx radio are for transmitting. Since each node (non BS) has just a single half-duplex radio, it can be either receiving or transmitting, but not doing both at the same time.  
Both experiments and large-scale simulations show high efficiency  of SNOW in latency and energy with a linear increase in throughput with the number of nodes, demonstrating its superiority  over existing designs~\cite{snow, snow2}.

%% file: system_model.tex
\section{System Model}\label{sec:system_model}

\begin{figure}[!htb]
\centering
\includegraphics[width=0.5\textwidth]{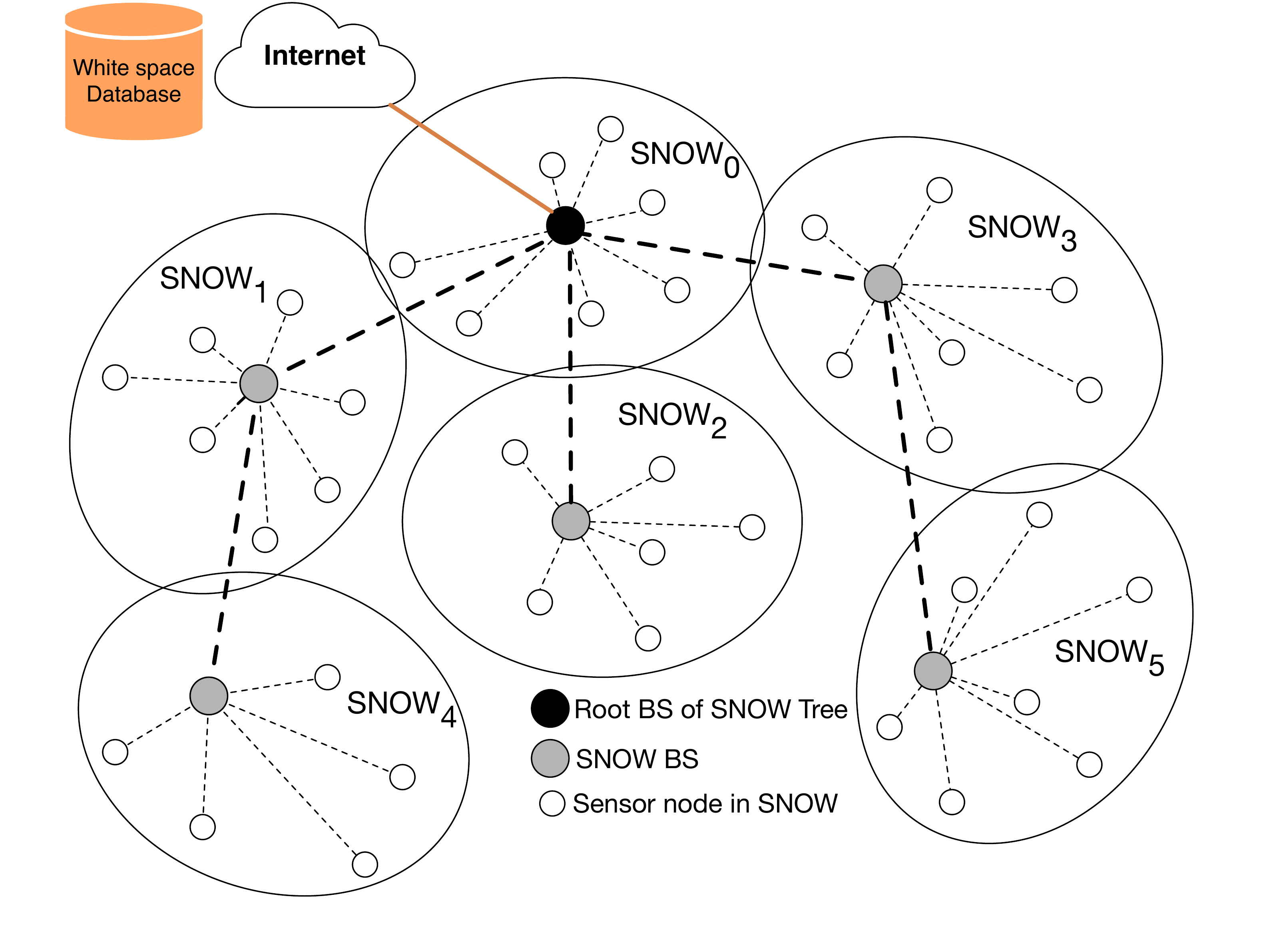}
\caption{A SNOW-tree.}
\label{fig:mesh}
\end{figure}
We consider  many coexisting SNOWs that are under the same management/control  and  need to coordinate among themselves for extended coverage in a  wide area or  to host different applications.   
As such, we consider an inter-SNOW network  as a SNOW-tree in the spirit of a  cluster tree  used in the new IEEE 802.15.4m standard~\cite{802154m}, each  cluster representing a personal area network under a coordinator. The {\bf root} of the tree is  connected to the white space database. In the similar spirit, our inter-SNOW network of the coordinated SNOWs is shown in Figure~\ref{fig:mesh} as a SNOW-tree. Each cluster is a star topology SNOW. All BSs form a tree that are connected through white space. Each BS is powerful or there can be multiple backup BSs for each cluster. So the chances of a BS failure is quite low in practice. Even if a BS fails, the root BS may reconstruct the tree.

Let there be a total of $N$ BSs (and hence $N$ SNOWs) in the SNOW-tree, denoted by BS$_0$, BS$_1$, $\cdots$, BS$_{N-1}$, where BS$_i$ is the base station of SNOW$_i$. BS$_0$ is the {\bf root BS} and is connected to the white space database through the Internet. The remaining BSs are in remote places where Internet connection many not be available. Those BSs thus depend on BS$_0$ for white space information.
Every BS is assumed to know the location of its operating area (its location and the locations of its nodes).  Localization is not the focus of our work and can be achieved through  manual configuration or some  existing WSN localization technique such as those based on ultrasonic sensors or other sensing modalities~\cite{snow}. 
BS$_0$  gets the location information of all BSs and  finds the white space channels for all SNOWs. It also knows the topology of the tree and allocates the spectrum among all SNOWs. Each BS  splits its assigned spectrum and assigns subcarriers to its nodes. For simplicity, we consider that all nodes in the tree transmit with the same transmission power and receive with the same receive sensitivity.

In an agricultural IoT, Internet connection is not available everywhere in the wide agricultural field. The farmer's home usually has the Internet connection and the root BS can be placed there. Microsoft's Farmbeats~\cite{vasisht2017farmbeats} project for agricultural IoT also exhibits such a scenario. Similarly, in a large oil field, the root BS can be in the office or control room. The considered SNOW-tree thus represents practical scenarios of wide area deployments in rural fields. The IEEE 802.15.4m standard also aims to utilize the white spaces under the exact same tree network model. We shall consider the scalability through a seamless integration and communication protocol among such coexisting SNOWs.

%% file: inter_snow_comm.tex
\section{Enabling Concurrent Inter-SNOW and intra-SNOW Communications}\label{sec:p2p}

Here we describe our inter-SNOW communication technique to enable seamless integration of the SNOWs for scalability. Specifically, we explain how we can enable concurrent inter-SNOW and intra-SNOW communications by exploiting the PHY design of SNOW.    To explain this we consider  \emph{peer-to-peer} inter-cluster communication in the SNOW-tree. That is, one node in a SNOW wants or needs to communicate with a node in another SNOW.
\begin{figure}[!htb]
\centering
\includegraphics[width=0.49\textwidth]{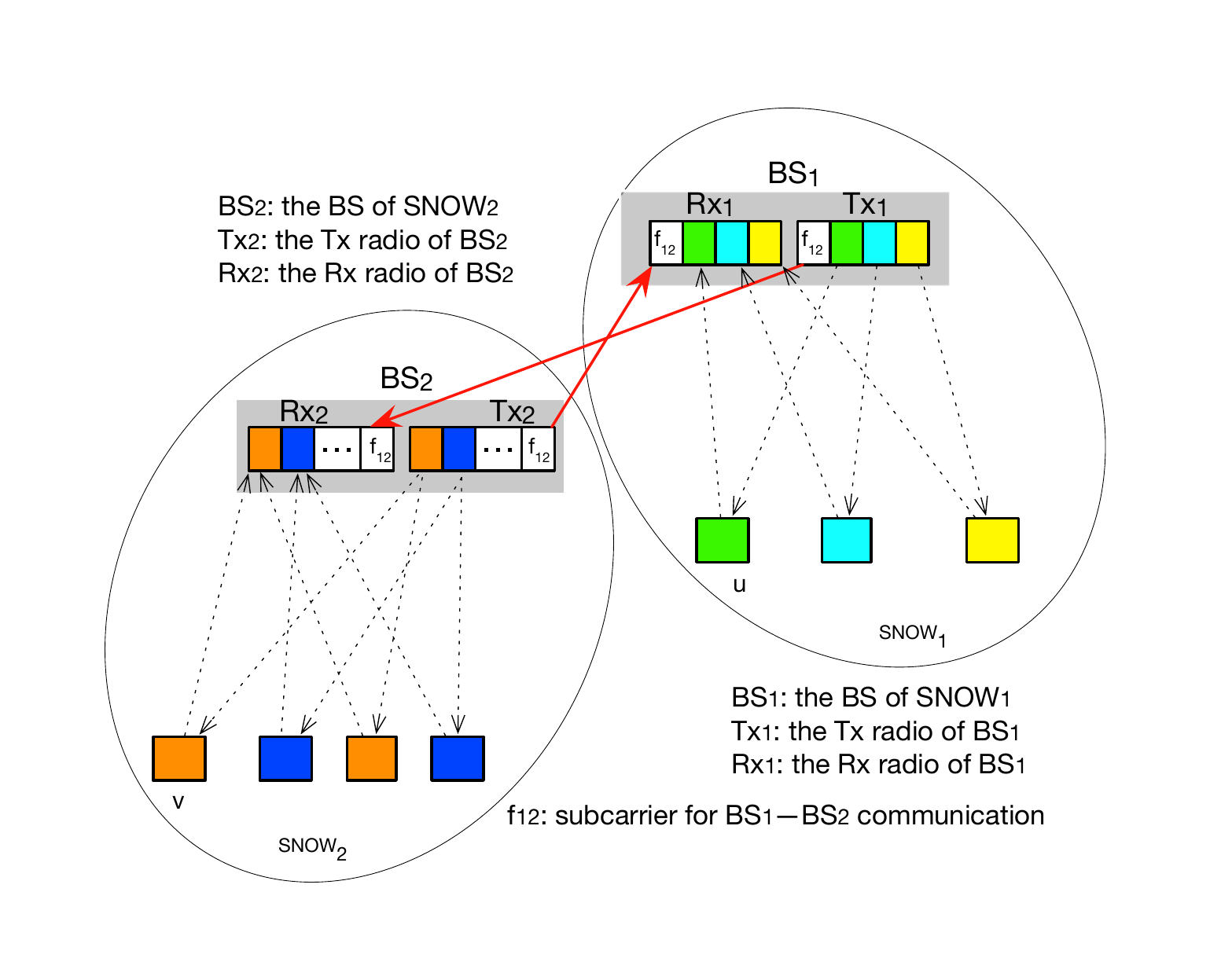}
\caption{Inter-SNOW communication.}
\label{fig:intercluster}
\end{figure}
  
For peer-to-peer communication across SNOWs, a node first sends its packet to its BS.
Note that two nodes may not communicate directly even if they are in communication range of each other as they may operate on different subcarriers.
The BS will then route to the destination SNOW's BS along the path given by the tree  which in turn  will  forward to the destination node.
Hence, the first {\bf question} is {\slshape ``How do two neighboring BSs exchange packets without interrupting their communication with their own nodes?"} Let us consider SNOW$_1$ and SNOW$_2$ as  two neighboring SNOWs in Figure~\ref{fig:intercluster} which will communicate with each other. We allocate a special subcarrier from both of their spectrum (i.e., a common subcarrier among the two BSs) that will be used for communication between these two BSs. 
For a tree link BS$_i\rightarrow$ BS$_j$, this subcarrier is denoted by $f_{i,j}$. To each tree link BS$_i\rightarrow$ BS$_j$, we assign a distinct $f_{i,j}$, eliminating interference among the BS transmissions made along the tree links. This is always feasible because the number ($N$) of SNOWs, and hence the number of tree links ($N-1$), is very small compared to the total number of subcarriers.
Additionally, if the connecting subcarrier that forms a tree link for BS-BS communication fails, another subcarrier is assigned since usually there is much overlap between two neighboring BSs.

As shown in Figure~\ref{fig:intercluster}, $f_{1,2}$ is a special subcarrier that enables BS$_1$-BS$_2$ communication as described above. D-OFDM allows us to encode any data on any subcarrier while the radio is transmitting. Thus the SNOW PHY will allow us to encode any time on any number of subcarriers and transmit. Exploiting this important feature of the SNOW PHY, Tx$_1$ radio will encode the packet on the subcarrier $f_{1,2}$ which is used for BS$_1$--BS$_2$ communication in Figure~\ref{fig:intercluster}.
If there are pending ACKs for its own nodes, they can also be encoded in their respective subcarriers. Then Tx$_1$ radio makes a single transmission. Rx$_2$ will receive it on subcarrier $f_{1,2}$ while the nodes of SNOW$_1$ will receive on their designated subcarriers. BS$_2$ can receive from BS$_1$ in the same way. 
They can similarly forward to next neighboring SNOWs. Thus both inter-SNOW and intra-SNOW communications can happen in parallel. 
Following are the several issues and our techniques to address those to enable such communication.

\subsection{Handling Collision in BS-BS Communication}
Using one subcarrier for  BS$_1$--BS$_2$ communication,   BS$_1$ and BS$_2$ cannot simultaneously transmit to each other. When Tx$_1$ transmits on  $f_{1,2}$, there is high energy on $f_{1,2}$ at Rx$_1$. The similar is the case  when Tx$_2$ transmits. If they start transmitting simultaneously, both packets will be lost. A straightforward solution is to use two different subcarriers for  Tx$_1\rightarrow$ Rx$_2$ and Tx$_2\rightarrow$ Rx$_1$ transmission. However, using two subcarriers dedicated for this may result in their underutilization and hinder  scalability. Hence, we  use a single subcarrier for BS$_1$--BS$_2$ communication and adopt random back-off within a fixed interval rule for this special subcarrier. That is, if BS-BS communication collides, they make random back-off after which they retry transmission.

\subsection{Dealing with Sleep/Wake up} 
When a node $u$ from SNOW$_1$ wants to send a packet to a node $v$ in SNOW$_2$, it first makes the transmission to BS$_1$ which then sends to BS$_2$ (Figure~\ref{fig:intercluster}). 
When BS$_2$ attempts to transmit to $v$, it can be sleeping and  BS$_2$ may be unaware of that.
To handle this, we adopt a periodic beacon that the BS of each SNOW sends to its nodes. The nodes are aware of the period of beacon. All nodes in a BS that are participating in peer-to-peer communication wake up for beacon. Thus, $v$ will wake up for beacon as it participates in peer-to-peer communication. BS$_2$ will 
encode $v$'s message on the subcarrier used by $v$ in the beacon. Thus, $v$ can receive the message from the beacon of BS$_2$.

%% file: technical.tex

\section{Handling Tradeoffs between Scalability and Inter-SNOW Interference}\label{sec:technical}
Our objective of integrating multiple SNOWs is scalability which can be achieved if every SNOW can support a large number of nodes. The number of nodes supported by a SNOW increases if the number of subcarriers used in that SNOW increases. However, if each SNOW uses the entire  spectrum available at its location,  there will be much spectrum overlap with the neighboring SNOWs. This will ultimately increase inter-SNOW interference, resulting in a lot of back-offs by the nodes during packet transmission. Like any other LPWAN, SNOW nodes are energy-constrained and cannot afford any sophisticated MAC protocol to avoid such interference, thereby wasting energy.
On the other end, if all neighboring SNOWs use non-overlapping spectrum, inter-SNOW interference will be minimized, but each SNOW in this way can support only a handful of nodes, thus degrading  the scalability. This tradeoff between scalability and inter-SNOW interference due to integration raises a spectrum allocation which cannot be solved using traditional spectrum allocation approach in wireless networks. We propose to accomplish such an allocation by formulating a {\bf\slshape Scalability Optimization Problem (SOP)} where our objective is to optimize scalability while limiting the interference.
To our knowledge, this problem is unique and never arose in other wireless domains. We now formulate SOP, prove its NP-hardness, and provide polynomial-time near-optimal solutions.

\subsection{SOP Formulation}\label{sec:prob_formulation}
The root BS knows the topology of the BS connections, accesses the white space database for each BS, and allocates the spectrum among the BSs. The spectrum allocation has to balance between scalability and inter-SNOW interference as described above. For SOP, we consider a uniform bandwidth $\omega$ of a subcarrier across all SNOWs. Let $Z_i$ be the set of  orthogonal subcarriers available at BS$_i$ considering $\alpha$ as the {\slshape fraction of overlap} between two neighboring subcarriers, where $0\le \alpha\le 0.5$ (as we found in our experiments~\cite{snow, snow2} that two orthogonal subcarriers can overlap at most up to half). Thus, if $W_i$ is the total available bandwidth at BS$_i$, then its total number of orthogonal subcarriers is given by $|Z_i| = \frac{W_i}{\omega\alpha} - 1.$

We consider that the values of $\omega$ and $\alpha$ are uniform across all BSs. Let the set of subcarriers to be assigned to BS$_i$ be $X_i\subseteq Z_i$,   with $|X_i|$ being the number of subcarriers in $X_i$.  We can consider the {\slshape total number of subcarriers}, $\sum_{i = 0}^{N-1} |X_i|$, assigned to all SNOWs as the  {\bf {\slshape scalability} metric}. We will maximize this metric. Every BS$_i$ (i.e., SNOW$_i$) requires a {\slshape minimum number of subcarriers} $\sigma_i$ to support its  nodes. Hence, we define {\bf Constraint } (\ref{c:const1})   to indicate the minimum and maximum number of subcarriers for each BS. If some communication in SNOW$_i$ is interfered by another communication in SNOW$_j$, then SNOW$_j$ is its {\bf \slshape interferer.} Since the root BS knows the locations of all BSs (all SNOWs) in the SNOW-tree, it can determine all interference relationships (which SNOW is an interferer of which SNOWs) among the SNOWs based on the nodes' communication range. 
\begin{figure*}[!htb]
\centering
\includegraphics[width=0.75\textwidth]{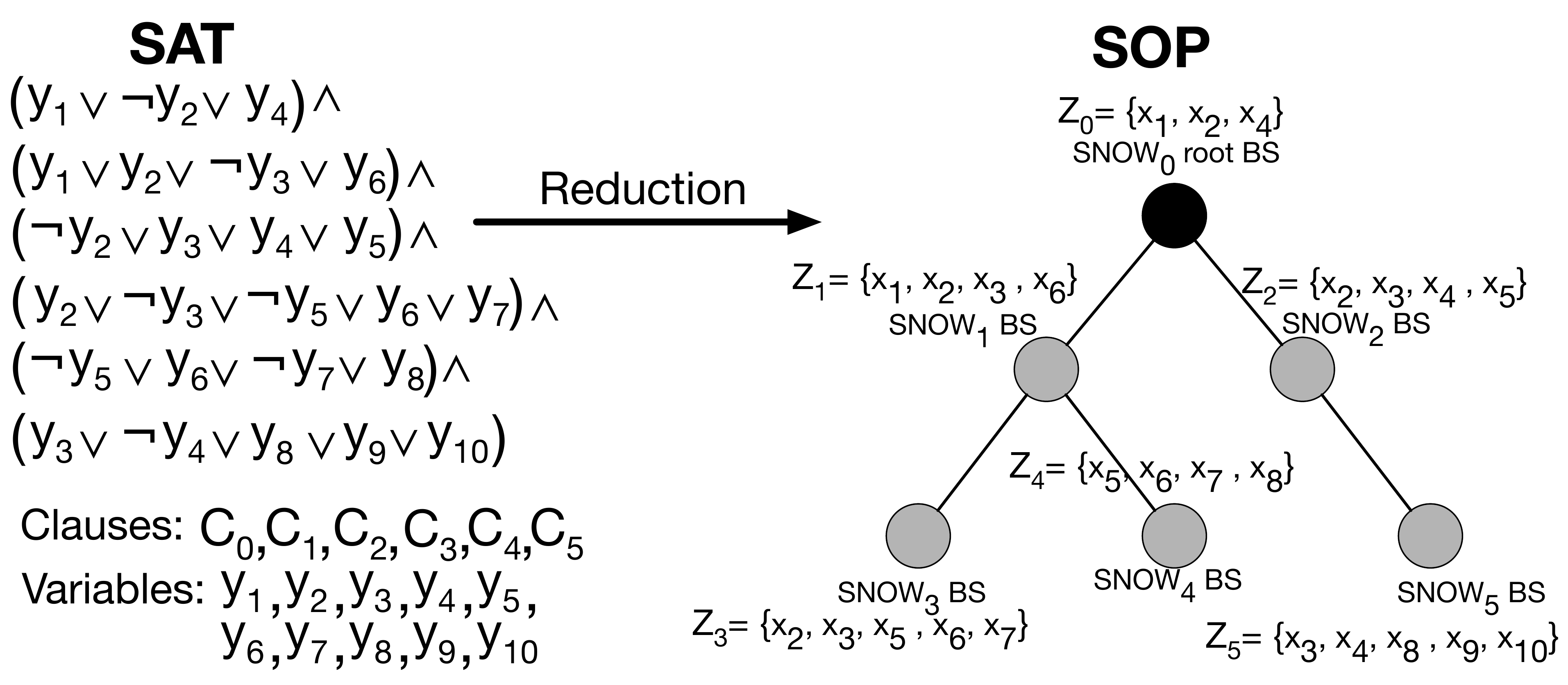}
\caption{Reduction from SAT.}
\label{fig:reduction}
\end{figure*}

Let $I_i\subset \{0, 1, \cdots, N-1\}$ be such that each SNOW$_j$ with $j\in I_i$ is an interferer of SNOW$_i$ (i.e., BS$_i$). In the SNOW-tree, let $p(i) \in \{0, 1, \cdots, N-1\}$  be such that BS$_{p(i)}$ is the parent of BS$_i$ and $Ch_j\subset \{1, 2, \cdots, N-1\}$ be such that each BS$_j$ with $j\in Ch_i$ is a child of BS$_i$. The SNOWs associated with a BS's parent and children are its interferer already, i.e., $(\{p(i)\} \cup Ch_i)\subseteq I_i$. To  limit inter-SNOW interference,  let  $\phi_{i, j}$ be the {\slshape maximum allowable number of subcarriers} that can overlap between two interfering SNOWS,  SNOW$_i$ and SNOW$_j$. As explained in Section~\ref{sec:p2p}, there must be at least one subcarrier common between a BS and its parent which is defined in {\bf Constraint} (\ref{c:const2}).
Note that we can also use Constraint (\ref{c:const2}) to set $\phi_{i, p(i)}$ to indicate the number of on demand subcarriers between BSs BS$_i$ and BS$_{p(i)}$ in a SNOW-tree. Sometimes the demand can change and the root BS will re-run the SOP algorithm to take it into account.
{\bf Constraint} (\ref{c:const3}) indicates the minimum and maximum number of overlapping subcarriers between other interfering pairs. Thus, SOP is formulated as follows where the root BS allocates the spectrum among all BSs (i.e., assigns subcarriers $X_i\subseteq Z_i$ to SNOW$_i$) in order to
\begin{align}
\centering
\text {Maximize~~~}  & \sum_{i = 0}^{N-1} |X_i| \nonumber \\
 \text{subject to~~~}& \sigma_i \leq |X_i| \leq |Z_i|, \; X_i \subseteq Z_i \label{c:const1} \\
                     & 1 \leq |X_i \cap X_{p(i)}| \leq \phi_{i, p(i)},  1 \le i < N  \label{c:const2}\\
                     & 0 \leq |X_i \cap X_j| \leq \phi_{i,j}, 0 \le i < N  \nonumber \\
                     &~~~~~~~~~~~~~~~~~~~~\forall j \in I_i  -  (\{p(i)\} \cup Ch_i) \label{c:const3}
\end{align}

SOP is a unique problem that we have observed first in integrating SNOWs. It is quite different from spectrum allocation in cellular network where towers are connected through a wired network and spectrum availability/dynamics~\cite{ismail2018low} do not change. Due to technology-specific features and unique communication primitive of SNOW, traditional channel allocation techniques for wireless networks (see survey~\cite{channel_wireless}), WSN (see survey~\cite{channel_wsn}), or cognitive radio networks (see survey~\cite{cognitive_channel_survey})  are also not applicable as SOP involves assigning a large number of subcarriers to each BS allowing some degree of overlaps among interfering BSs for enhanced scalability. In the following, we will first characterize SOP and then propose its solution strategy.

\subsection{NP-Hardness of SOP}

We now prove that SOP is NP-hard which can be proved through a reduction from the {\slshape SAT (Boolean Satisfiability)} problem. The SAT problem asks whether there exists a truth assignment that makes all clauses true~\cite{morgado2013iterative}. Theorem~\ref{thm:np_hard_proof} formally proves the NP-hardness of SOP by proving that its decision version is NP-complete.

\begin{theorem}\label{thm:np_hard_proof}
Given a SOP for SNOW-tree, it is NP-complete to decide whether it is feasible or not.
\end{theorem}

\begin{IEEEproof} Given an instance of SOP in SNOW-tree with overlapping spectrum assignment for $N$ BSs and $m$ subcarriers, where BS$_i$, $0 \le i < N$ gets $m_i$ number of subcarriers. It is verifiable in $O(Nm)$ time whether the subcarrier assignment is feasible or not. Hence, the problem is in NP. To prove NP-hardness, we reduce an arbitrary instance $\mathcal{I}(SAT)$ of SAT to an instance $\mathcal{I}(SOP)$ of the SOP in SNOW-tree and show that $\mathcal{I}(SAT)$ has an \emph{interpretation} that satisfies a \emph{boolean formula} if and only if $\mathcal{I}(SOP)$ is feasible.

Let $\mathcal{I}(SAT)$ have $m$ boolean variables $y_0, y_1, y_2, ..., y_{m-1}$ and $N$ clauses $C_0, C_1, \cdots, C_{N-1}$ in conjunctive normal form. Now, for the set of variables in $\mathcal{I}(SAT)$ we create a set of subcarriers $Z = \{x_0, x_1, \cdots, x_{m-1}\}$ in $\mathcal{I}(SOP)$ that are available in SNOW-tree. Then, we create one SNOW BS$_i$ in $\mathcal{I}(SOP)$ for each clause $C_i$ in $\mathcal{I}(SAT)$. Also, we create one subset $Z_i \in Z$ for each BS$_i$ that corresponds to subset of boolean variables in clause $C_i$. 
 As an example, consider a boolean formula $(y_1 \lor \neg y_2 \lor y_4) \land (y_1 \lor y_2 \lor \neg y_3 \lor y_6) \land (\neg y_2 \lor y_3 \lor y_4 \lor y_5) \land (y_2 \lor \neg y_3 \lor \neg y_5 \lor y_6 \lor y_7) \land (\neg y_5 \lor y_6 \lor \neg y_7 \lor y_8) \land (y_3 \lor \neg y_4 \lor y_8 \lor y_9 \lor y_{10})$ of 10 variables and 6 clauses in $\mathcal{I}(SAT)$, thus in $\mathcal{I}(SOP)$, $Z_0 = \{x_1, x_2, x_4\},\; Z_1 = \{x_1, x_2, x_3, x_6 \},\; Z_2 = \{x_2, x_3, x_4, x_5 \}, \cdots,$ and $Z_{5} = \{x_3, x_4, x_8, x_9, x_{10}\}$. If a boolean variable $y_k$ exists as a positive literal in clause $C_i$ and negative literal in $C_j$, then corresponding BS$_i$ (i.e. SNOW$_i$) and BS$_j$ (i.e. SNOW$_j$) interfere each other and $x_k \in \{Z_i \cap Z_j\}$ is the interfering subcarrier between them. Thus, setting $y_k$ to $true$ in $\mathcal{I}(SAT)$ will yield assigning subcarrier $x_k$ to BS$_i$ or BS$_j$, and vice versa. In the previous example, if $y_2$ is set to $true$, then BS$_1$ and BS$_3$ get subcarrier $x_2$ and not BS$_0$ and BS$_2$.

To build the SNOW-tree, we consider BS$_0$ as the root BS that corresponds to clause $C_0$. 
We draw an edge between BS$_i$ and BS$_j$ if corresponding clauses $C_i$ and $C_j$ have at least one common positive or negative literal. The number of such literals in $\mathcal{I}(SAT)$ represents $\phi^{}_{ij}$ in $\mathcal{I}(SOP)$. While creating the SNOW-tree, we \emph{do not} draw an edge between BS$_j$ and BS$_k$ if BS$_j$ $\in (\{p(i)\} \cup Ch_i)$ and BS$_k$ $\in (\{p(i)\} \cup Ch_i)$, where, $i \ne j \ne k$. Thus, no loops are created and the number of edges in SNOW-tree become $N-1$, as shown in Figure~\ref{fig:reduction}. The whole reduction process runs in $O(m^2\lg N)$ time.

Suppose that $\mathcal{I}(SAT)$ has an interpretation that satisfies the boolean formula. Thus, each clause $C_i$ is also $true$. Also, a subset of variables in each clause $C_i$ is $true$ that corresponds to the subset of subcarriers $X_i$ that is assigned to BS$_i$ in $\mathcal{I}(SOP)$. The number of variables in clause $C_i$ that are set to $true$ represents the minimum number of subcarriers $\phi_i$ in $\mathcal{I}(SOP)$. Also, no two interfering BS$_i$ and BS$_j$ get more than $\phi^{}_{ij}$ number of common subcarriers between them. We also include a common subcarrier between neighboring BS$_i$ and BS$_j$ if there is none already, thus considering corresponding literal in $\mathcal{I}(SAT)$ as $true$ which does not change the satisfiability of boolean formula. Such inclusion also does not violate right hand side condition of Constraints (\ref{c:const2}) and (\ref{c:const3}). Thus, $\mathcal{I}(SOP)$ has a feasible subcarrier assignment where the root BS assigns at least $N$ subcarriers in total to all the BSs SNOW-tree, each having at least one.

Now, let $\mathcal{I}(SOP)$ have a feasible subcarrier assignment in SNOW-tree. Thus the root BS assigns at least $N$ subcarriers to $N$ BSs, each having at least one. Since each BS$_i$ in $\mathcal{I}(SOP)$ represents a clause $C_i$ in $\mathcal{I}(SAT)$ and two neighboring BS$_i$ and BS$_j$ in $\mathcal{I}(SOP)$ have at least one common subcarrier and SNOW-tree is connected, each clause $C_i$ has at least one literal that is set to be $true$. Thus, we have an interpretation in $\mathcal{I}(SAT)$ that satisfies the boolean formula.
\end{IEEEproof}

\subsection{Efficient Greedy Heuristic for SOP}\label{sec:proposed_solution}

Since an optimal solution of SOP cannot be achieved in polynomial time unless P=NP, we first propose an intuitive and efficient polynomial-time greedy heuristic. 
In the beginning, our greedy heuristic assigns to each BS the entire spectrum available in its location. Our heuristic then keeps removing subcarriers from all the BSs until all the constrains of SOP are satisfied. Our goal here is to remove as less subcarriers as possible from each BS to maximize the {\slshape scalability} metric.
\begin{algorithm}
\DontPrintSemicolon
\KwData{ $Z_i$ for BS$_i$, $0 \le i < N$ in a SOP Instance.}
\KwResult{Subcarriers $X_i$ for BS$_i$, $0\le i <N$.}
\BlankLine

\For{$\text{each } BS_i \text{ in } \text{the SNOW-tree}$}{
$X_i = Z_i$.
}

\For{$\text{each } BS_i \text{ in } \text{inter-SNOW Tree}$}{
   \For{$\text{each } BS_j\; \in\; I_i$}{
        Let, $Z_{i,j} = Z_i \cap Z_j$.
        
        \For{$\text{each subcarrier}\; x_l\; \in\; Z_{i,j}$}{
            \If{$|X_i \cap X_j| > \phi_{i,j}$}{
                \If{$|X_i| \ge |X_j| \; and\; |X_i| > \sigma_i$ }{
                
                    Delete $x_l$ from $X_i$.
                }
                \ElseIf {$|X_j| > \sigma_j$}{
                
                    Delete $x_l$ from $X_j$.
                }
                \Else (\tcc*[f]{Infeasible solution}){
                
                    Don't delete $x_l$ from $X_i$ or $X_j$. 
                }
            }
            \Else{
                Break.
            }
        }
    }
}

\caption{Greedy Heuristic Algorithm}
\label{algo:greedy}
\end{algorithm}

Our greedy heuristic is described as follows. The root BS first greedily assigns to BS$_i$ all the subcarriers that are available at the location of BS$_i$ (i.e.,  the entire spectrum available in BS$_i$'s  location).  Note that such an assignment maximizes the  {\slshape scalability} metric $\sum_{i = 0}^{N-1} |X_i|$, but violates the constrains of SOP. Specifically, it satisfies Constraint (\ref{c:const1}), but may violate Constraints (\ref{c:const2}) and (\ref{c:const3}) that are defined to keep the BSs connected as a tree and to limit interference between neighboring or interfering BSs by limiting their common usable subcarriers.  Now, with a view to satisfying those two constrains, the heuristic greedily removes some subcarriers  that are common between interfering BSs. Such removal of subcarriers is done to make the least decrease in the scalability and to ensure that Constraint (\ref{c:const1}) is not violated.  In other words, it tries to keep the subcarrier assignment balanced between BSs. Specifically, for every interfering BS pair, BS$_i$ and BS$_j$, we do the following until they satisfy Constraints (\ref{c:const2}) and (\ref{c:const3}):  Find the next common subcarrier between them and remove it from BS$_i$ if $| X_i | > | X_j |$ and  $| X_i | > \sigma_i$; otherwise remove it from BS$_j$ if $| X_j | > \sigma_j$.

The pseudocode of our greedy heuristic is shown as Algorithm~\ref{algo:greedy}. As shown in the pseudo code, the heuristic may not find feasible solution in some rare cases where some BS pairs, BS$_i$ and BS$_j$,  cannot satisfy the condition $|X_i \cap X_j| \le  \phi_{i,j}$.  In such cases, we can either use the infeasible solution and use  the found subcarrier allocation or relax the constraints for those BSs (violating the constraints) by changing their values of $\sigma_i$ or $\phi_{i,j}$ in Constraints (\ref{c:const1}),  (\ref{c:const2}), and (\ref{c:const3}) of the SOP. 

\subsubsection{Time Complexity of the Greedy Heuristic} 
Theorem~\ref{thm:greedy} provides the time complexity of Algorithm~\ref{algo:greedy}.

\begin{theorem}\label{thm:greedy}
Algorithm~\ref{algo:greedy} has a time complexity of $O(N^2M\lg M)$, where $N$ is the number of SNOWs and $M= \max \{   | Z_i|  \mid  0\le i <N\}$.
\end{theorem}
\begin{IEEEproof}
Since the SNOW-tree has $N$ base stations (or  $N$ SNOWs),  Algorithm~\ref{algo:greedy} will find intersection of the subcarriers for each of  $O(N^2)$ pairs of BSs (line 5 of Algorithm~\ref{algo:greedy}). 
Finding intersection of the subcarriers for a pair of BSs takes   $O(M\lg M)$ time, where $M= \max \{   | Z_i|  \mid  0\le i <N\}$. Thus, the time complexity of Algorithm~\ref{algo:greedy} is  $O(N^2M\lg M)$. 
\end{IEEEproof}

\subsection{Approximation Algorithm for SOP}\label{sec:bounded_algorithm}

While the heuristic (Algorithm~\ref{algo:greedy}) can be highly efficient
in practice, we also propose an algorithm for which we can derive an analytical performance bound. 
Our reduction used in Theorem~\ref{thm:np_hard_proof} provides the key insights for developing such an approximation algorithm. 
Our key observation from the reduction is that a solution approach for SOP can be developed by extending a solution for 
the MAX-SAT (Maximum Satisfiability) problem and by incorporating the constraints of the former. MAX-SAT, a generalized version of SAT,
asks to determine the maximum number of clauses, of a given
Boolean formula in conjunctive normal form, that can be made
true by an assignment of truth values to the variables of the
formula~\cite{maxsat}. The observation allows us to leverage the well-established results for MAX-SAT. 
Specifically, we leverage a very simple but  analytically efficient approach adopted for MAX-SAT solution, and incorporate the SOP constraints 
to develop a constant approximation algorithm for SOP.

\begin{algorithm}
\DontPrintSemicolon
\KwData{ $Z_i$ for BS$_i$, $0 \le i < N$ in a SOP Instance.}
\KwResult{Subcarriers $X_i$ for BS$_i$, $0\le i <N$.}
\BlankLine

\For{$\text{each } BS_i \text{ in } \text{the SNOW-tree}$}{
$X'_i = X''_i =\emptyset;$ 
}

Let, $Z = Z_0 \cup Z_2 \cup \cdots \cup Z_{N-1}$

\For (\tcc*[f]{step 1}) {$\text{each subcarrier} \;x_l \in Z$}{
        Uniformly and independently add $x_l$ with a probability of $\frac{1}{2}$ to $X'_i$, $\forall i: x_l \in Z_i$.
}

  \If (\tcc*[f]{violates Constraint \ref{c:const1}}) {$\exists i $ such that $ |X'_i| <  \sigma_i $}{ 
Let, $Z^{\prime} = (Z_0 - X'_0) \cup (Z_1 - X'_1) \cup \cdots \cup (Z_{N-1} - X'_{N-1})$

\For (\tcc*[f]{step 2})  {$\text{each subcarrier} \;x_k \in Z^{\prime}$}{
    Uniformly and independently add $x_k$ with a probability of $\frac{1}{2}$ to $X''_i$, $\forall i: x_k \in (Z_i - X'_i)$.
}

}     

\For{$\text{each } BS_i \text{ in } \text{the SNOW-tree}$}{
$X_i = X'_i  \cup X''_i;$.
}

\caption{Probabilistic 1/2-Approximation Algorithm}
\label{algo:approx}
\end{algorithm}

Considering $\Omega$ as the total weight of all clauses, a simple approximation algorithm for MAX-SAT sets each variable to true with probability $\frac{1}{2}$. 
By linearity of expectation, the expected weight of the satisfied clauses is at least $\frac{1}{2}\Omega$, thus making the approach a randomized $\frac{1}{2}$-approximation algorithm. 
In solving the SOP  in a similar spirit, we shall consider assigning a {\slshape subcarrier} to a {\slshape SNOW} in place of a {\slshape variable} to a {\slshape clause}. Choosing a probability other than $\frac{1}{2}$ would require us to calculate different probabilities for different subcarriers based on the level of interference they contribute to different BSs which involves a costly approach. Therefore, it is very difficult and impractical for us to develop a faster approximation algorithm based on our proposed approach. Since MAX-SAT  does not have  Constraints (\ref{c:const1}), (\ref{c:const2}), and (\ref{c:const3}), we modify such a probabilistic assignment whose  pseudocode is shown  as Algorithm~\ref{algo:approx} to take into account these constraints.

Algorithm~\ref{algo:approx} assigns subcarriers to the BSs in two steps. In step 1, it assigns each distinct subcarrier $x_l$ in the SNOW-tree uniformly and independently with probability of $\frac{1}{2}$ to each BS$_i$ such that $x_l \in Z_i$ (i.e., the BS where the subcarrier is available). The set of subcarriers that BS$_i$ gets after this step is $X'_i$. Thus, the expected number of subcarriers assigned to BS$_i$ in this step is $E[ |X'_i|]= \frac{|Z_i|}{2}$. Similarly, the expected number of common subcarriers between two interfering BSs, BS$_i$ and BS$_j$, after step 1 is 
$E[ |X'_i  \cap  X'_j   |]= \frac{|Z_i  \cap  Z_j |}{4}$. Our experiments (Sec. \ref{sec:experiments}, \ref{sec:simulations}) show that two interfering BSs can use even up to 60\% of their total available common subcarriers. That is, the values $\phi_{i,j}$ in Constraints  (\ref{c:const2}) and (\ref{c:const3}) can be up to 60\%  of  $|Z_i  \cap  Z_j |$. Thus after step 1, the probability of satisfying Constraints  (\ref{c:const2}) and (\ref{c:const3})  is very high. 
Hence, if some BS$_i$ violates Constraint  (\ref{c:const1}), i.e., if $|X'_i|  < \sigma_i$, we repeat subcarrier assignment in the same way in step 2. 
Specifically, step 2 assigns each distinct subcarrier $x_k$ uniformly and independently with probability of $\frac{1}{2}$ to each BS$_i$ such that $x_k \in Z_i  -  X'_i $. If $X''_i$  is the set of subcarriers assigned to BS$_i$ in step 2, then BS$_i$ finally gets subcarriers $X_i = X'_i  \cup X''_i$. While step 2  increases the probability of satisfying Constraints  (\ref{c:const1}), it decreases that of satisfying the other two constraints which was very high before this step. Hence, we do not adopt any further subcarrier addition.


\subsubsection{Performance Analysis}

As described above, Algorithm~\ref{algo:approx} sometimes can end up with an infeasible solution for SOP.  However, as we describe below, such chances are quite low, and the probability of finding a feasible solution is quite high ($\approx 1$) (Lemma~\ref{thm:prob}). Then Theorem~\ref{thm:bound} proves that the Algorithm has an approximation ratio of $\frac{1}{2}$ for any solution it provides (feasible or infeasible).  

\begin{lemma}\label{thm:prob}
The probability of satisfying all the constraints of SOP is $\approx 1$.
\end{lemma}

\begin{IEEEproof}
As described before, after step 1  of Algorithm~\ref{algo:approx},   $E[|X_i^{\prime}|] = \frac{|Z_i|}{2}$ for each BS$_i$; and $E[|X_i^{\prime} \cap X_j^{\prime}|] = \frac{|Z_i \cap Z_j|}{4}$, for each interfering BS pairs,  BS$_i$ and BS$_j$.  Step 2 runs  only if Constraint (\ref{c:const1}) remains violated after step 1. Thus, if step 2 does not run, the probability of satisfying  Constraint (\ref{c:const1}) is 1. Similarly, if step 2 runs, $E[|X_i^{\prime\prime}|] = \frac{|Z_i - X_i^{\prime}|}{2}$ and $E[|X_i^{\prime\prime} \cap X_j^{\prime\prime}|] = \frac{|(Z_i - X_i^{\prime}) \cap (Z_j - X_j^{\prime})|}{4}$. Now, if both steps run, the expected number of subcarriers assigned to BS$_i$ is

\begin{align} 
\centering
    E[|X_i|] &= E[|X'_i|] +  E[|X''_i|]\nonumber \\ 
    &=  \frac{|Z_i|}{2} + \frac{|Z_i - X'_i|}{2} \nonumber \\
    &=     \frac{|Z_i|}{2} + \frac{|Z_i|} {2}  -  \frac{|Z_i|} {4} = \frac{3}{4} |Z_i|. 
\end{align}

Note that the value of $\sigma_i$ is set usually much smaller than the above value as a BS does not want to use all available subcarriers, allowing other SNOWs to use those. 
Thus, the probability of satisfying Constraint (\ref{c:const1}) is $\approx 1$.

If step 2 does not run, then the expected number of common subcarriers between each interfering BS pairs, BS$_i$ and BS$_j$, is $E[|X_i \cap X_j|]  = \frac{|Z_i \cap Z_j|}{4}$. As we have discussed before, the value of $\phi_{i,j}$ in Constraints  (\ref{c:const2}) and (\ref{c:const3})  is usually above $\frac{|Z_i  \cap  Z_j |}{2}$, which is twice the value of  $E[|X_i \cap X_j|]$. Thus, the probability of  satisfying Constraints (\ref{c:const2}) and (\ref{c:const3}) is also $\approx 1$.  If step 2 runs, then

\begin{align}
\centering
    E[|X_i \cap X_j|] &= E[|X_i^{\prime} \cap X_j^{\prime}|] +  E[|X_i^{\prime\prime} \cap X_j^{\prime\prime}|] \nonumber \\
    &= \frac{|Z_i \cap Z_j|}{4} +  \frac{|(Z_i - X_i^{\prime}) \cap (Z_j - X_j^{\prime})|}{4} \nonumber \\
    &= \frac{|Z_i \cap Z_j|}{4} + \frac{| Z_i \cap Z_j |}{4}  -  \frac{| X'_i \cap X'_j |}{4} \nonumber \\
     &= \frac{|Z_i \cap Z_j|}{4} + \frac{| Z_i \cap Z_j |}{4}  -  \frac{| Z_i \cap Z_j |}{16} \nonumber \\
     &= \frac{ 7}{16} |Z_i \cap Z_j|     <     \frac{|Z_i  \cap  Z_j |}{2}
\end{align}

which means that  the probability of satisfying Constraints (\ref{c:const2}) and (\ref{c:const3}) is $\approx 1$ even if step 2 runs. Thus, the probability of satisfying all constraints is $\approx 1$.
\end{IEEEproof}

\begin{theorem}\label{thm:bound}
 Algorithm~\ref{algo:approx} has an approximation ratio of $\frac{1}{2}$. 
\end{theorem}

\begin{IEEEproof}
Since an optimal value of the objective (scalability metric) is unknown, a conservative upperbound is given by $OPT = \sum_{i=0}^{N-1} |Z_i| $. 
If step 2 of the algorithm does not run, according to probabilistic assignments of subcarriers in step 1 of Algorithm~\ref{algo:approx}, we have
in step 1,

\begin{align}
\centering
E[\text{total }X_i] &=\sum_{i=0}^{N-1}\sum_{l = 0}^{|Z_i|-1}|x_l|\Pr \{ x_l\text{  assigned to }BS_i\} \nonumber \\
&= \sum_{i=0}^{N-1}|Z_i|.\frac{1}{2} \nonumber \\
&= \frac{1}{2} \sum_{i=0}^{N-1} |Z_i|    \ge \frac{1}{2}OPT
\end{align}

If step 2 of Algorithm~\ref{algo:approx} runs,

\begin{align}
\centering
E[\text{total }X''_i] &= \frac{1}{2} \sum_{i=0}^{N-1} |Z_i  -      X'_i| \nonumber \\
&=\frac{1}{2} \sum_{i=0}^{N-1}|Z_i|    -   \frac{1}{2} \sum_{i=0}^{N-1}|X'_i| \nonumber \\
&=\frac{1}{2} \sum_{i=0}^{N-1}|Z_i|    -   \frac{1}{4} \sum_{i=0}^{N-1}|Z_i| \nonumber \\
&=  \frac{1}{4} \sum_{i=0}^{N-1}|Z_i|.
\end{align}

\vspace{1mm}
Now using linearity of expectation, if step 2 runs,

\begin{align}
\centering
E[\text{total }X_i] &=  \frac{1}{2} \sum_{i=0}^{N-1} |Z_i|   +   \frac{1}{4} \sum_{i=0}^{N-1} |Z_i| \nonumber \\
                           & =   \frac{3}{4} \sum_{i=0}^{N-1} |Z_i|    \ge \frac{3}{4}OPT
\end{align}

Thus the approximation bound follows. 
\end{IEEEproof}

\subsubsection{Time Complexity of the Approximation Algorithm}
Theorem~\ref{thm:approx} provides the time complexity of Algorithm~\ref{algo:approx}.

\begin{theorem}\label{thm:approx}
Algorithm~\ref{algo:approx} has a time complexity of $O(NM\lg M)$, where $N$ is the number of SNOWs and $M= \max \{   | Z_i|  \mid  0\le i <N\}$.
\end{theorem}

\begin{IEEEproof}
Since the SNOW-tree has $N$ BSs, Algorithm~\ref{algo:approx} will do  {\slshape union} operations of all the $O(N)$ BSs' subcarriers (in line 3) in the step 1. If step 2 runs, {\slshape subtraction} and {\slshape union} operations will be done.  {\slshape Unions} and {\slshape subtractions} will run in $O(M\lg M)$ time, where $M= \max \{   | Z_i|  \mid  0\le i <N  \} $. Thus, the time complexity of Algorithm~\ref{algo:approx} is $O(NM\lg M)$.
\end{IEEEproof}

As we shall describe in Sections~\ref{sec:experiments} and \ref{sec:simulations} through evaluations, our heuristic (Algorithm~\ref{algo:greedy}) performs better in terms of scalability, energy consumption, and  latency while Algorithm~\ref{algo:approx} provides theoretical performance guarantee.

%% file: experiment.tex
\section{Implementation}\label{sec:implementation}

We implement our proposed SNOW technologies on the GNU Radio~\cite{gnuradio} platform using USRP devices that can operate between 70MHz - 6GHz of spectrum~\cite{usrp}. We have 9 USRP (2 B210, 4 B200, and 3 USRP1) devices. 
To demonstrate the effectiveness of SOP in intra-SNOW communication we use 2x2 devices in 2 different SNOW BSs (each having one Tx-Radio and one Rx-Radio), where one BS is assigned 3 nodes (3 USRPs) and the other BS is assigned 2 nodes (2 USRPs). On the other hand, to demonstrate the inter-SNOW communication, we use 2x3 devices in 3 different SNOW BSs (each having one Tx-Radio and one Rx-Radio). In this case, each BS is assigned one USRP device as node.
\begin{figure}[!htb]
\centering
\includegraphics[width=0.45\textwidth]{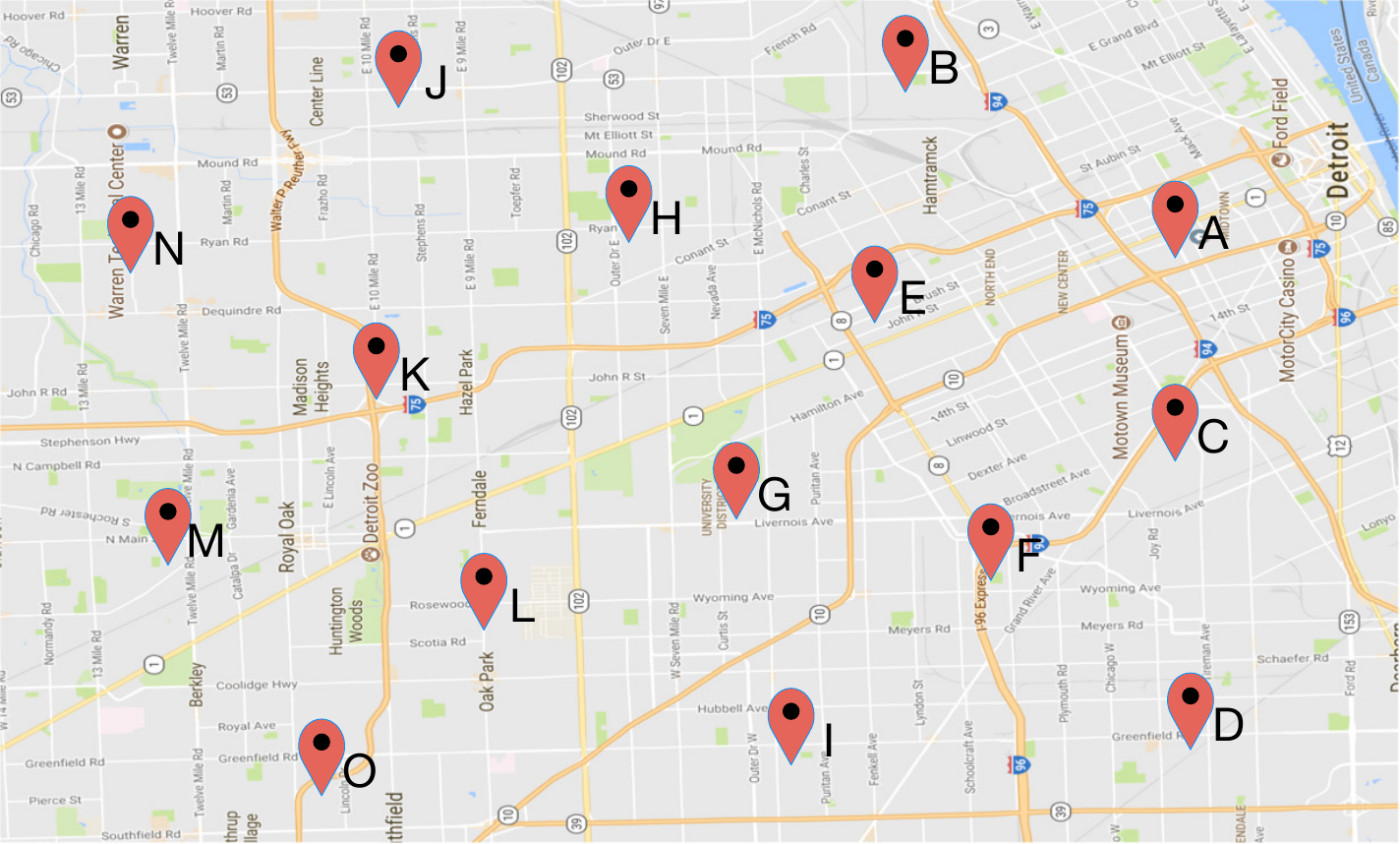}
\caption{SNOW BS positions used in experiments (and simulations).}
\label{fig:testbed}
\end{figure}

We evaluate the performance of our design by experimenting at 15 different candidate locations covering approximately (25x15)km$^2$ of a large metropolitan area in the city of Detroit, Michigan (Figure~\ref{fig:testbed}). 
Due to our limited number of USRP devices (3 BSs each having one node to demonstrate inter-SNOW communication) in real experiments, we create 5 different SNOW-trees at different candidate locations and do the experiments separately. 
In experiments, we choose to create 3 SNOWs to demonstrate the integration of as many SNOWs as we can with our limited number of devices, and most importantly to cover more area using a SNOW-tree. 
In~\cite{snow, snow2}, we have already performed extensive experiments considering multiple nodes in a single SNOW. Hence, here we will show the intra-SNOW communication using 2 SNOW BSs one having 3 nodes and the other having 2 nodes.
However, later in simulations, we create a single SNOW-tree of 15  SNOWs each having 1000 nodes. We perform experiments on white space availability at different locations and determine the values of $\phi_{i, p(i)}$ and $\phi_{i, j}$ in Constraints (\ref{c:const2}) and (\ref{c:const3}), respectively. We compare the performance of our greedy heuristic and our approximation algorithm for SOP with a direct allocation scheme. A \emph{direct allocation scheme} is unaware of scalability and inter-SNOW interference and hence will assign each BS all the subcarriers that are available at its location. Moreover, we perform exhaustive experiments on both intra- and inter-SNOW communications.

\section{Evaluation}\label{sec:eval}
In this section, we evaluate the performance of our SOP algorithms in inter- and intra-SNOW communications through experiments and simulations.
\subsection{Experiments}\label{sec:experiments}
\subsubsection{Experimental Setup}\label{sec:exp_setup} 

Our testbed location has white spaces ranging between 518 and 686MHz (TV channels 21--51) for different BSs. We set each subcarrier bandwidth to 400kHz which is the default subcarrier bandwidth in SNOW~\cite{snow, snow2}. We use 40-byte (including header, random payload, and CRC) packets with a spreading factor of 8, modulated or demodulated as BPSK (Binary Phase-Shift Keying). With the similar spirit of IEEE 802.15.4, we set the Tx power to 0dBm in the SNOW nodes for energy efficiency. Receive sensitivity is set to -94dBm both in SNOW BSs and the nodes. Meanwhile, BSs transmit with a Tx power of 15dBm ($\approx$40mW) to their nodes and neighboring BSs that is the maximum allowable Tx-power limit in most of the white space channels at our testbed location. For energy calculations at the nodes, we use the energy profile of TI CC1070 RF unit by Texas Instruments that can operate in white spaces~\cite{cc1070}. Unless stated otherwise, these are our default parameter settings.

\subsubsection{Finding Allowable Overlap of Spectrum}\label{sec:exp_whitespaces}

\begin{figure}[!htb] 
    \centering
      \subfigure[Available white spaces (denoted as TV channel indices used in the US) at different BS locations. A dot in the figure means that the TV channel in x-axis is white space at the location in y-axis.\label{fig:channels}]{
    \includegraphics[width=0.32\textwidth]{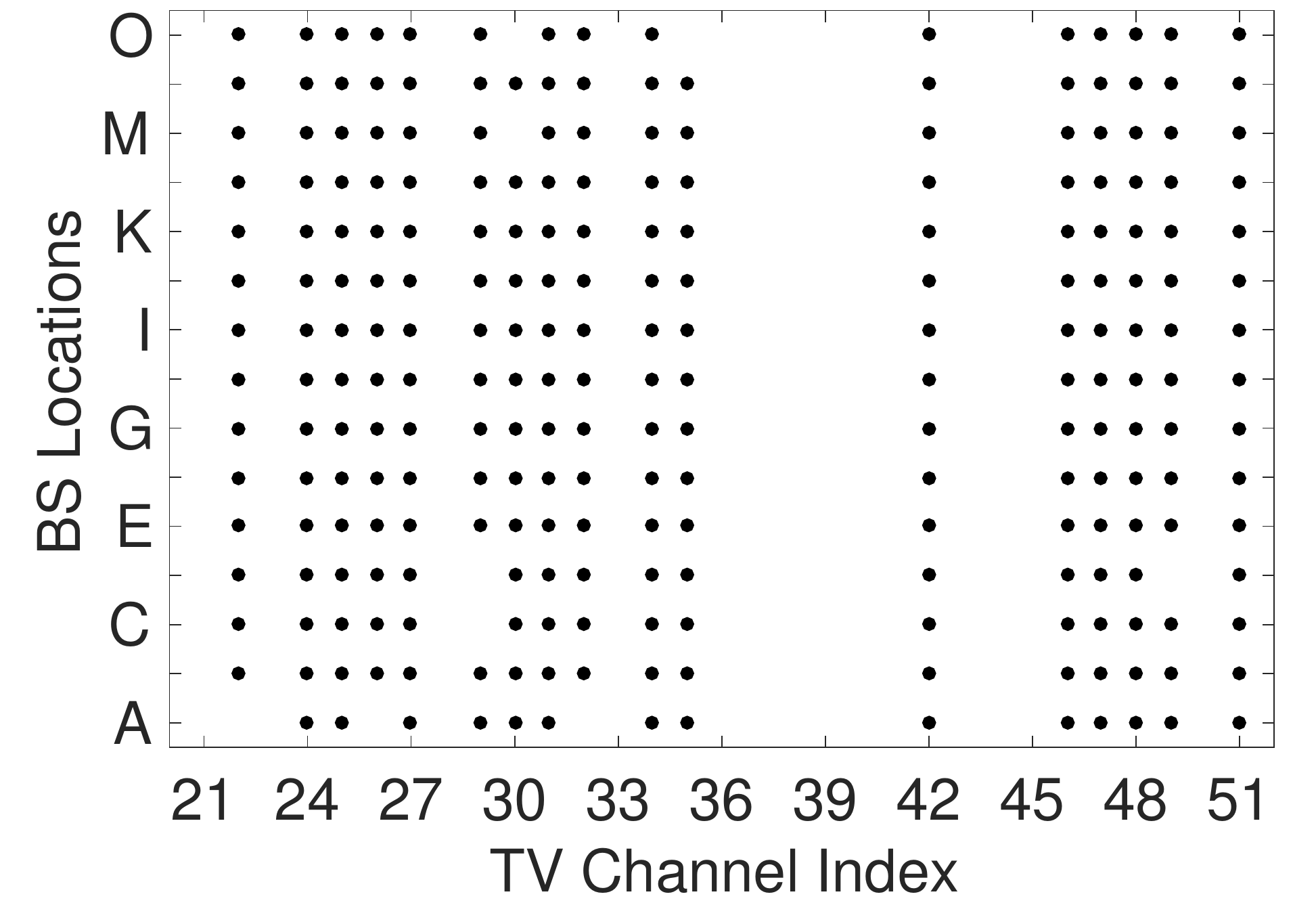}
      } \vspace{-0.1in}
    \vfill
      \subfigure[Reliability in communication with different magnitude of overlaps in white spaces between BSs in different SNOW-trees.\label{fig:overlaps}]{
        \includegraphics[width=.32\textwidth]{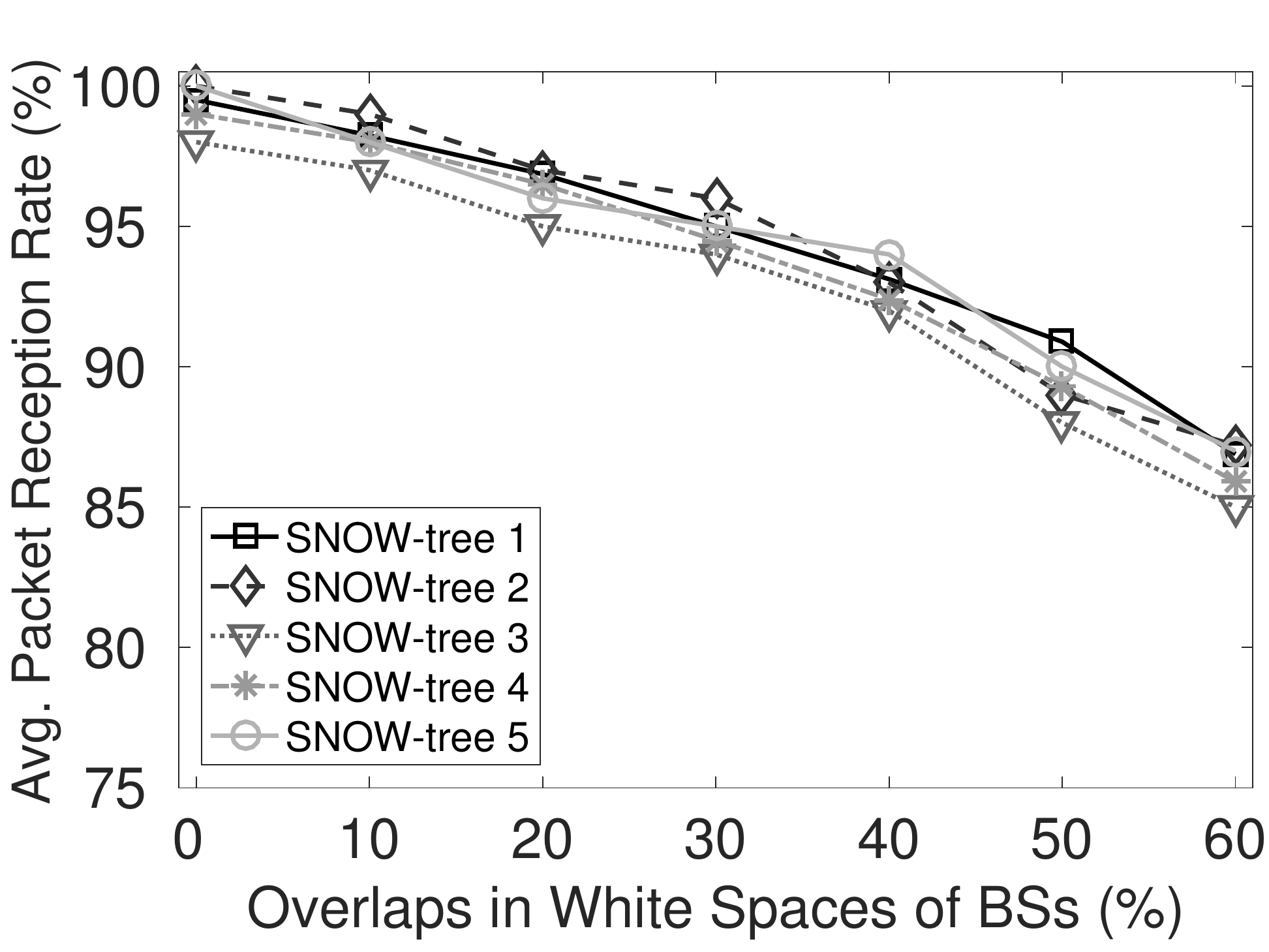}
      }
    \caption{White spaces availability and reliability in different SNOW-trees.}
    \label{fig:primitives}
\end{figure}

We first determine how many subcarriers can be common  between two interfering SNOWs without degrading their performance.  
We determine white spaces at 15 different locations from a cloud-hosted database~\cite{wdb}.
Figure~\ref{fig:channels} shows the available white spaces at different locations confirmed by both database and sensing. Also, we conduct experiments on 5 different SNOW-trees to determine the maximum allowable number of common subcarriers between interfering BSs. Locations of BSs in 5 trees are (1) B, A, E; (2) D, C, F; (3) G, I, L; (4) J, H, K; (5) N, M, O; respectively, where the BS in the middle location in each SNOW-tree is the root BS. In this paper, we also identify the SNOW BSs by their location indices. In each tree, we allow BSs to operate with different magnitudes of white space overlap between them. To determine the maximum allowable number of common subcarriers between interfering BSs in a tree, each node hops randomly to all the subcarriers that are available in its BS location and sends consecutive 100 packets to its BS. Each node repeats this procedure 1000 times.

As shown in Figure~\ref{fig:overlaps}, the BSs in each tree can overlap 60\% of their white spaces to yield an average Packet Reception Rate (PRR) of 85\%. We consider that an 85\% PRR is an acceptable rate. This figure also shows that the average PRR decrease with the increase in the magnitude of overlap. Finding the maximum allowable overlap needs to be done only once in the beginning of the network operation and may be recomputed if there is a significant change (e.g., some BS or a large number of nodes leave or join) in the network. A network deployment may choose its magnitude of overlap based on the target application’s quality of service (QoS) requirements. We thus set the values of $\phi_{i,p(i)}$ and $\phi_{i,j}$ in Constraints (\ref{c:const2}) and (\ref{c:const3}), respectively, based on this experiment. Finding the optimal values of these variables is out of the scope of this paper.

\subsubsection{Evaluating the Scalability Metric}\label{sec:exp_sop}
\begin{figure}[!htb]
    \centering
      \subfigure[Scalability metric values achieved in different SNOW-trees.\label{fig:subs}]{
        \includegraphics[width=.32\textwidth]{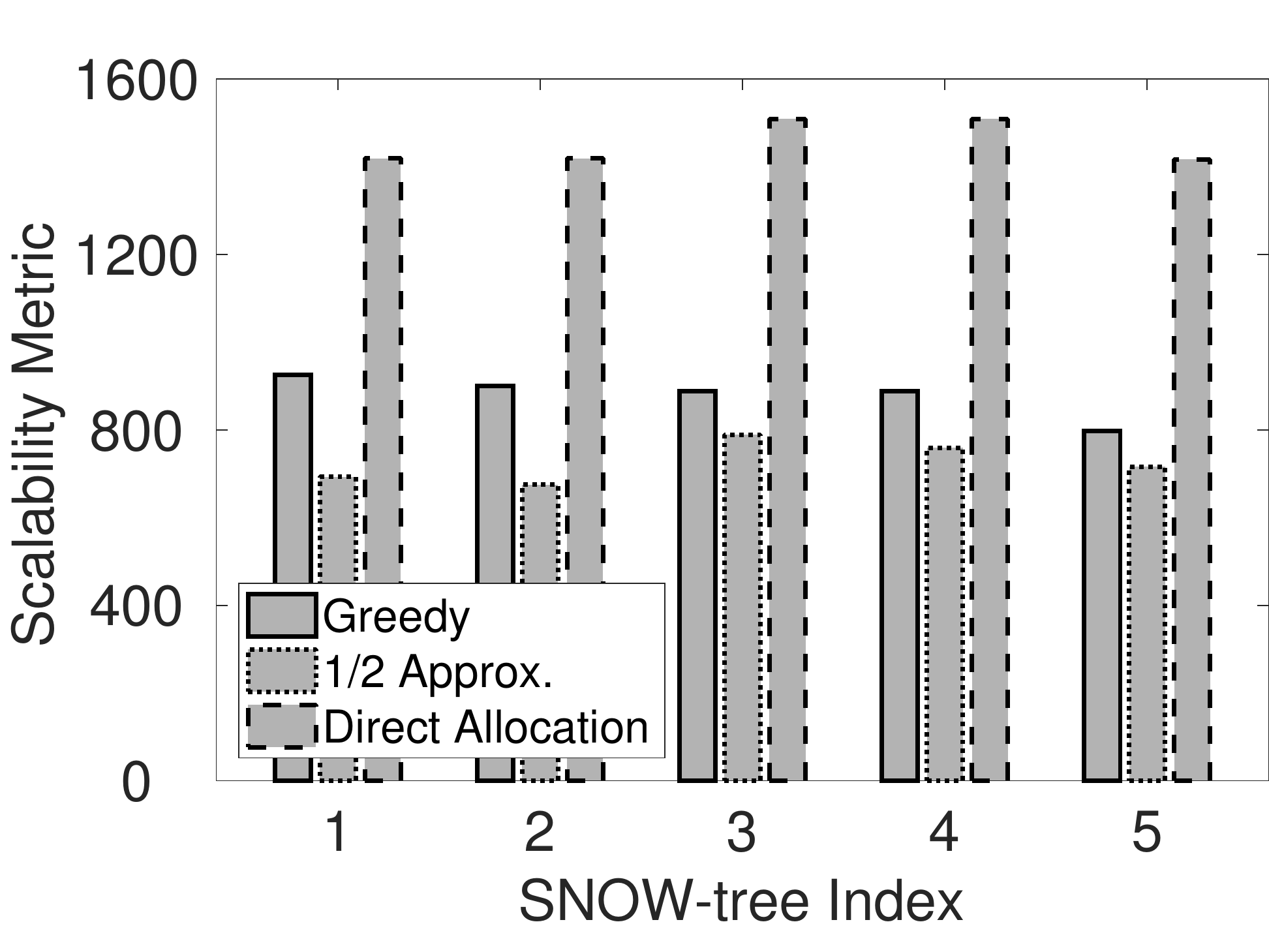}
      }
      \vfill
      \subfigure[Execution time of different SOP algorithms.\label{fig:latency_algo}]{
        \includegraphics[width=.32\textwidth]{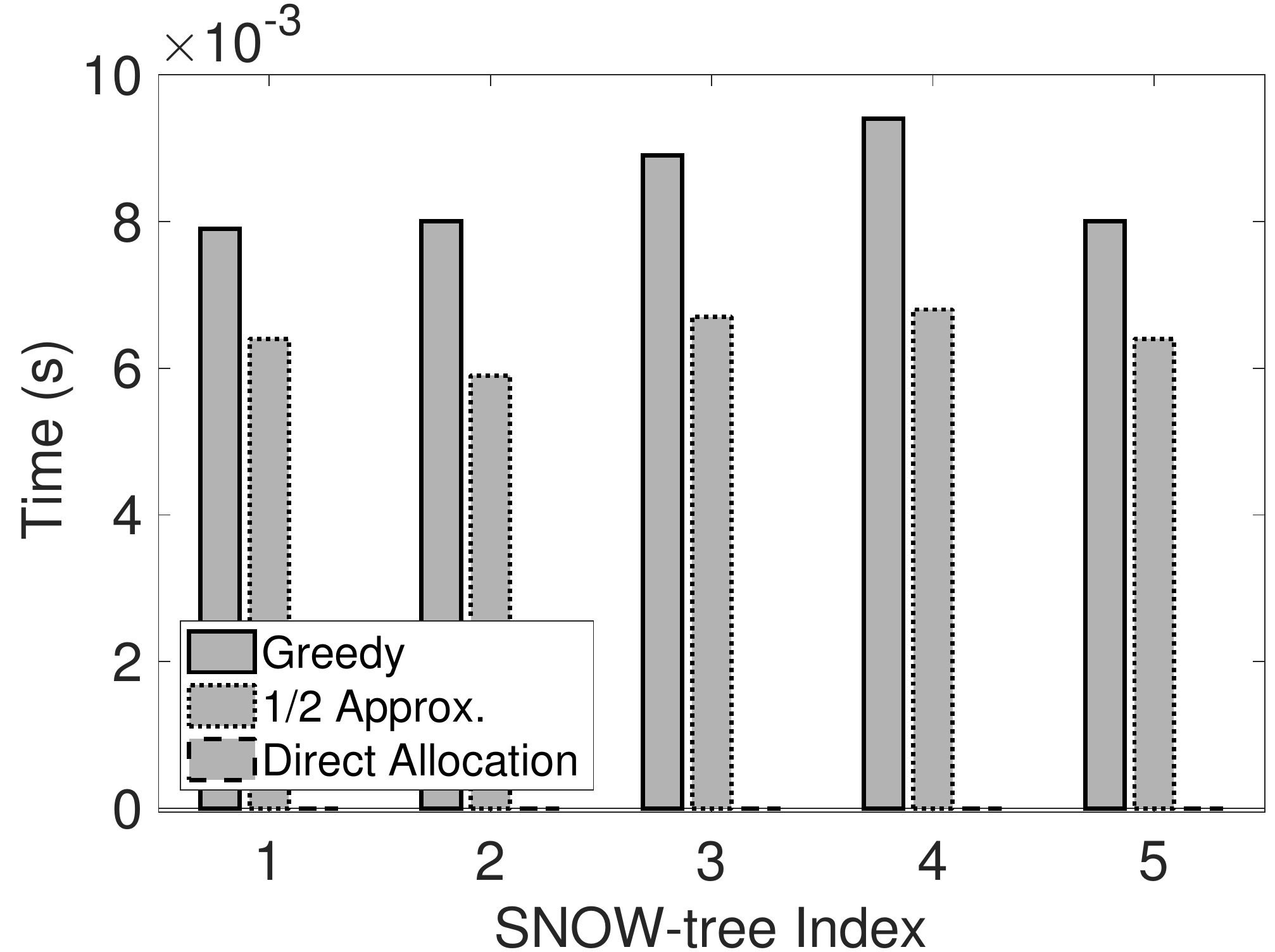}
      }
    \caption{Scalability metric and execution time of SOP algorithms by different root BSs in different SNOW-trees.}
    \label{fig:sop}
\end{figure}
\begin{figure*}[!htb]
    \centering
      \subfigure[Reliability in different SNOW BSs\label{fig:snow_reliability}]{
    \includegraphics[width=0.32\textwidth]{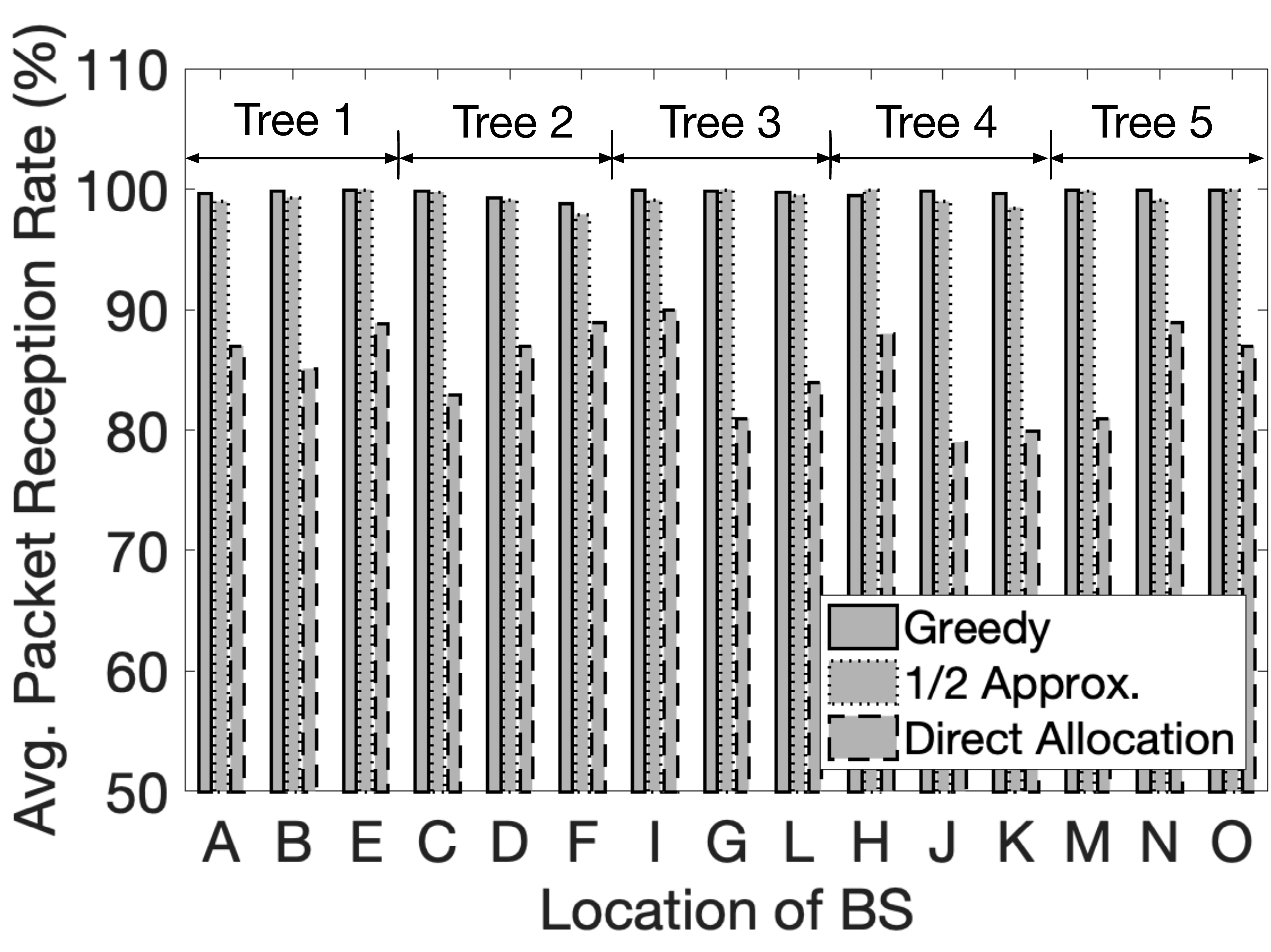}
      }
    \hfill
      \subfigure[Average Latency in intra-SNOW comm.\label{fig:snow_latency}]{
        \includegraphics[width=.31\textwidth]{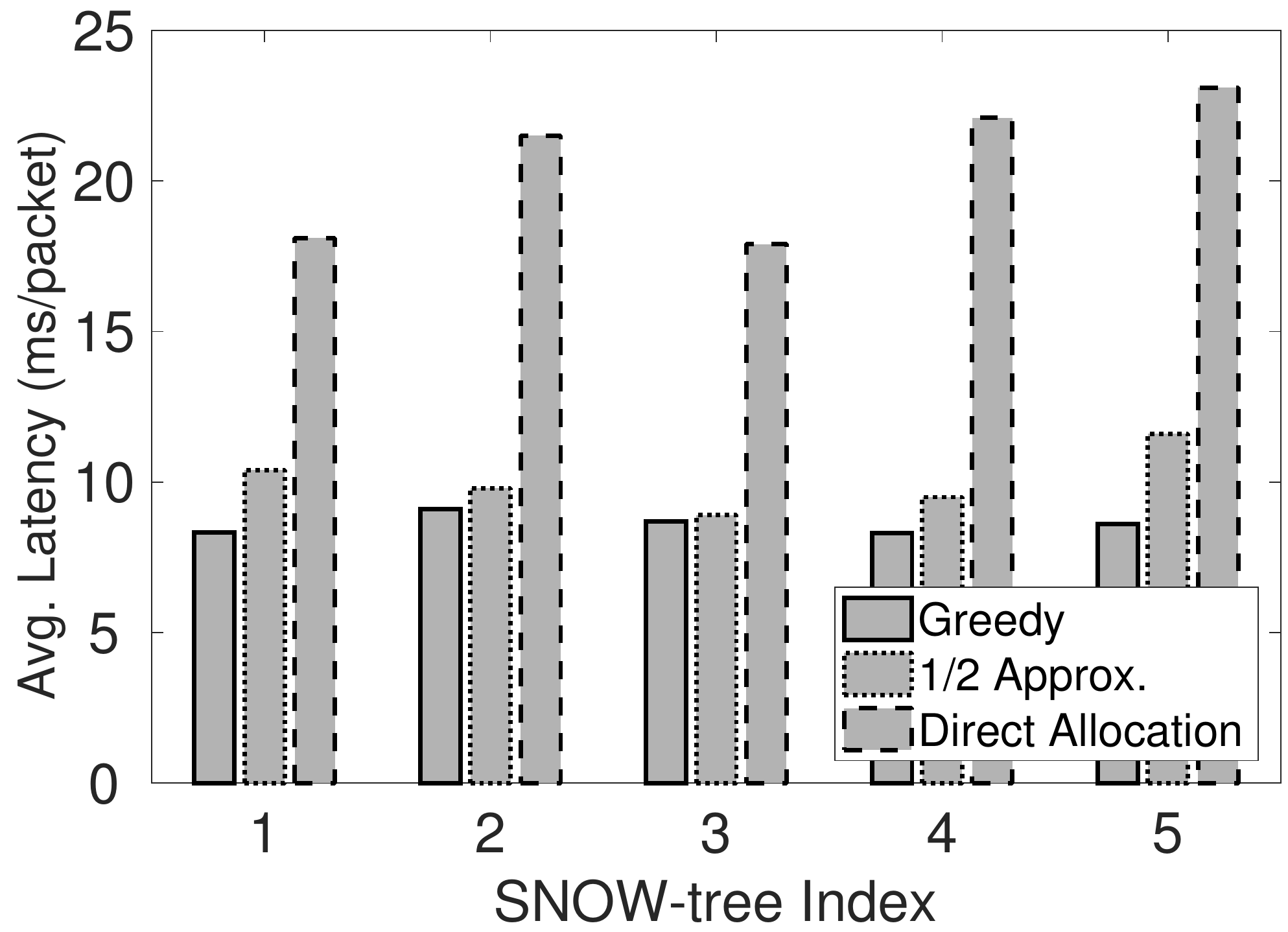}
      }
      \hfill
      \subfigure[Energy consumption in intra-SNOW comm.\label{fig:snow_energy}]{
        \includegraphics[width=.31\textwidth]{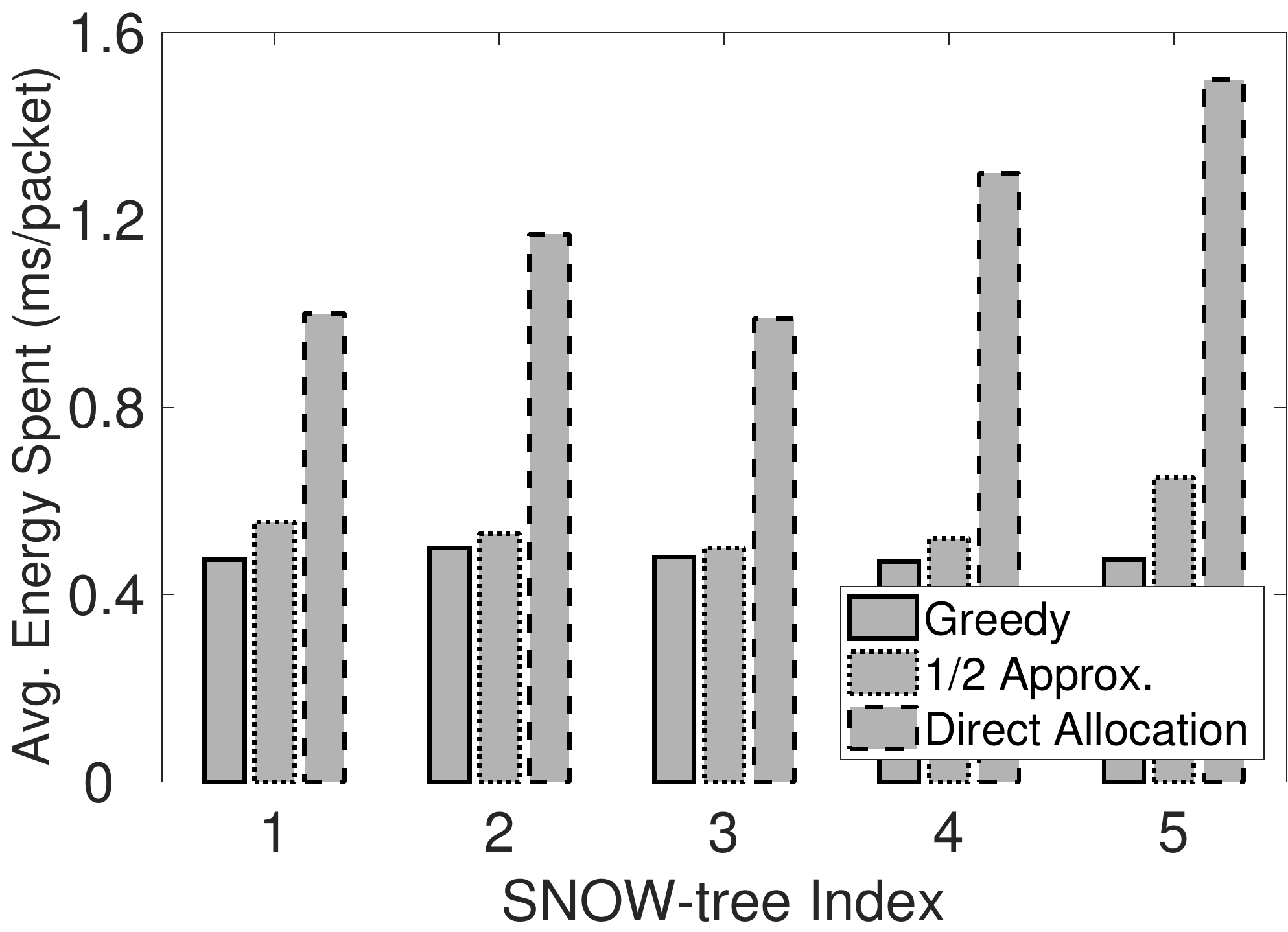}
      }
    \caption{Performance of intra-SNOW communications in different SNOW-trees.}
    \label{fig:snow_comm}
\end{figure*}
To demonstrate the performances in maximizing the scalability metric under our approaches and the baseline approach, we set the value of $\sigma_i$ in Constraint (\ref{c:const1}) to 100 for all the BSs. 
We choose the same value for each BS since most (13 out of 15) of the BS locations have the same set of white space channels. 
Figure~\ref{fig:subs} shows the values of the scalability metric achieved in 5 different SNOW-trees using our greedy heuristic,  approximation algorithm, and the direct allocation approach. 
This figure shows that the direct allocation scheme assigns more subcarriers to all BSs.  Our later experiments will show that such an assignment suffers in terms of reliability, latency, and energy consumptions compared to our greedy heuristic and approximation algorithm due to its violation of Constraints (\ref{c:const2}) and (\ref{c:const3}) of SOP. 
Also, our greedy heuristic can offer higher scalability than our approximation approach, while the latter can be preferred when analytical performance bound is a concern. Thus, our greedy heuristic can be more effective in practice (even though its performance bound was not derived).

Figure~\ref{fig:latency_algo} shows the time taken by our greedy heuristic, our approximation algorithm, and direct allocation scheme to assign subcarriers to BSs. Our greedy heuristic observes 0.094ms compared to 0.068ms for our approximation algorithm in worst case in SNOW-tree 4. In the figure, time taken by the direct allocation scheme is not visible as it is approximately 0ms (since it does not employ any intelligent technique). However, time taken by our greedy heuristic and our approximation algorithm are very low and practical.

\begin{figure*}[!htb]
    \centering
      \subfigure[Reliability in inter-SNOW communication\label{fig:p2p_reliability}]{
    \includegraphics[width=0.315\textwidth]{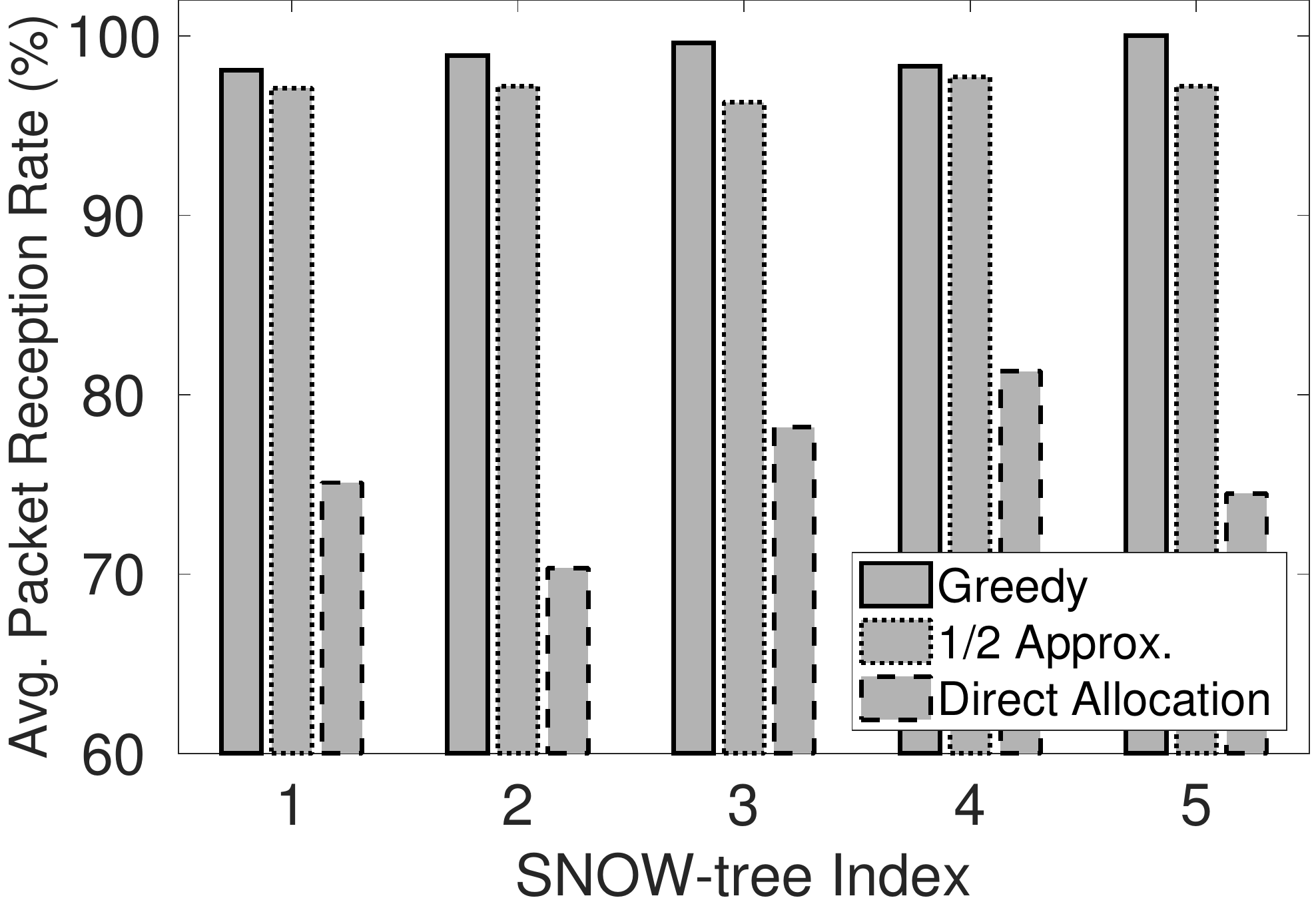}
      }
    \hfill
      \subfigure[Latency in inter-SNOW communication\label{fig:p2p_latency}]{
        \includegraphics[width=.315\textwidth]{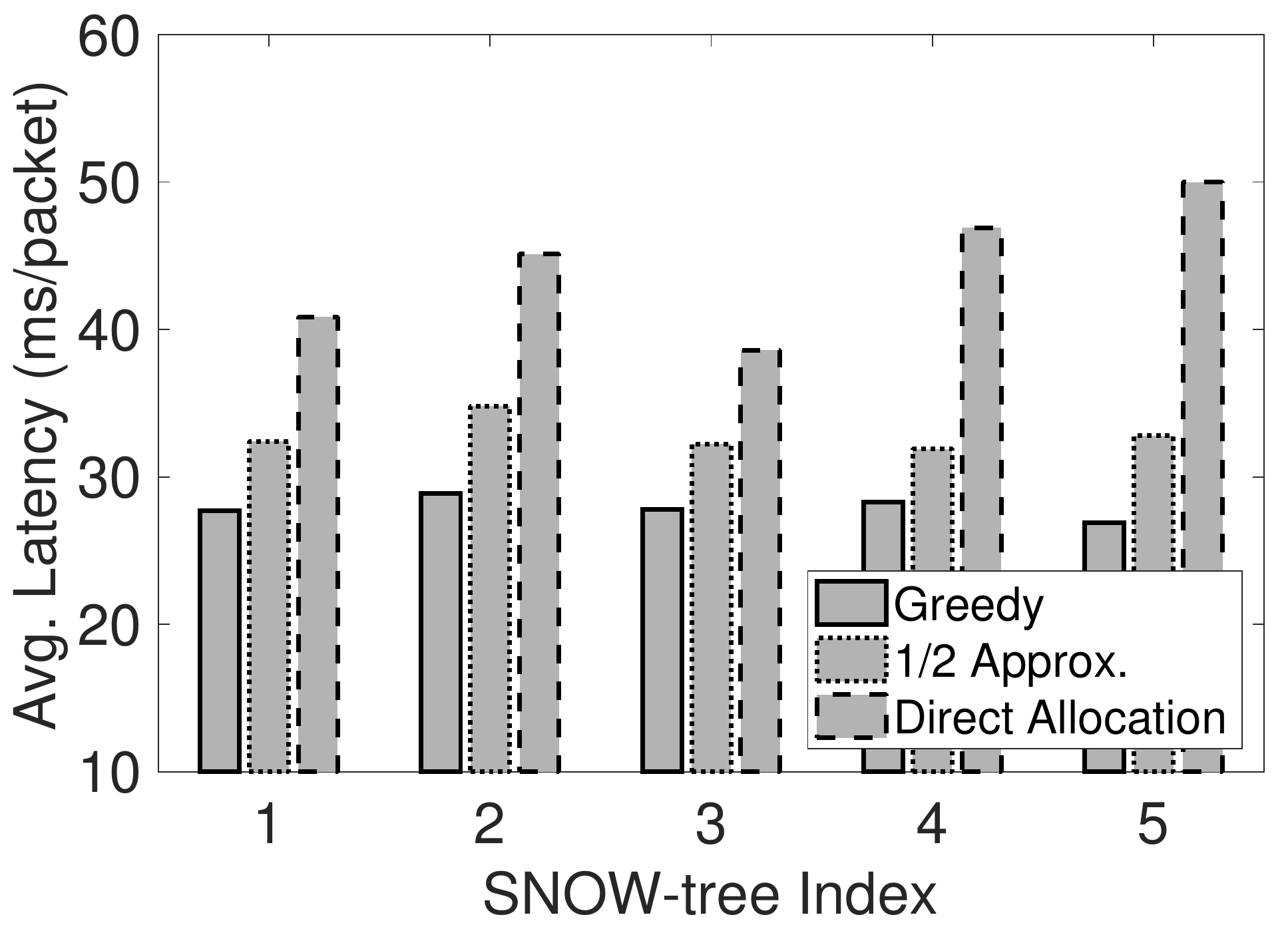}
      }
      \hfill
      \subfigure[Energy consumption in inter-SNOW comm.\label{fig:p2p_energy}]{
        \includegraphics[width=.315\textwidth]{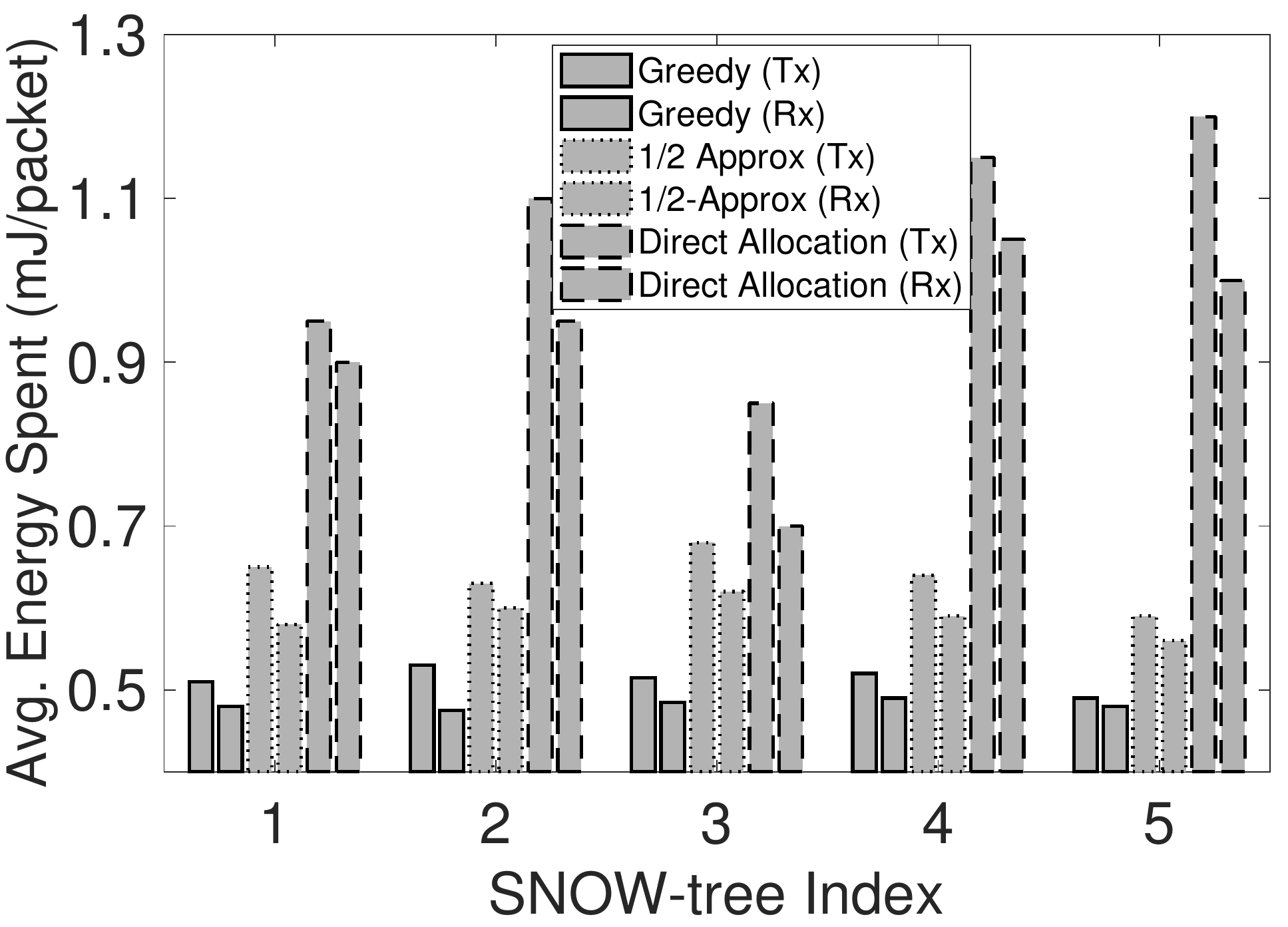}
      }
    \caption{Performance of inter-SNOW communications in different SNOW-trees.}
    \label{fig:p2p_comm}
\end{figure*}
\subsubsection{Experiments on Intra-SNOW Communication}
In this section, we demonstrate intra-SNOW communication performance when multiple interfering SNOWs are integrated together to coexist. Due to our limited number of USRP devices, we choose two interfering SNOWs in a SNOW-tree and run intra-SNOW communications independently at the same time. For example in SNOW-tree 1, SNOWs at locations A and B perform intra-SNOW communications. Here, SNOWs at locations A and B are assigned 3 and 2 nodes, respectively (as explained in Section~\ref{sec:implementation}). 
Similarly, we allow SNOWs at locations B and C; C and A to do the same, respectively. In experiments, each node under a SNOW hops randomly on different subcarriers assigned by our greedy heuristic algorithm and sends 100 consecutive packets to the BS. We repeat the same set of experiments when subcarrier assignment is done by our approximation algorithm and the direct allocation scheme. 
We allow the nodes in a SNOW to hop randomly across different subcarriers to emulate as if all the subcarriers of that SNOW were assigned to different nodes. Figure~\ref{fig:snow_comm} shows the reliability, latency, and energy consumption in intra-SNOW communication under different SOP algorithms.

Figure~\ref{fig:snow_reliability} shows the average PRR in different SNOW BSs. In each SNOW-tree, the average PRR at each SNOW BS is calculated from all 3 pairs of intra-SNOW communication experiments. The highest average PRR is approximately 100\% in SNOW BSs located at E, I, M, N, and O, while the lowest average PRR is approximately 98.9\% in SNOW BS located at F when the subcarriers assigned by our greedy heuristic algorithm is used. For our approximation algorithm, the highest and lowest average PRR values are approximately 100\% and 97.9\%, respectively. For the direct allocation scheme, these values are 89\% and 79\%, respectively.
Figure~\ref{fig:snow_latency} shows that the average latency to successfully deliver an intra-SNOW packet to a SNOW BS is lower in all SNOW-trees while the subcarriers assigned by our greedy heuristic algorithm are used. For example, the average latency per packet is as low as 8.3ms in SNOW-tree 4 compared to 9.5ms and 22.1ms for our approximation algorithm and direct allocation subcarrier assignments, respectively.
Figure~\ref{fig:snow_energy} shows that the average energy consumption for each packet is also lower in all SNOW-trees when our greedy subcarrier assignment is used. In SNOW-tree 4, the average energy consumption per packet is as low as 0.47mJ compared to 0.52mJ and 1.31mJ for approximation and direct allocation subcarrier assignments, respectively. 
Thus, all the experiments in Figure~\ref{fig:snow_comm} confirm that both our greedy heuristic and approximation algorithm are practical choices for SOP.
\begin{figure*}[!htb]
    \centering \vspace{-0.15in}
      \subfigure[Locations of SNOW BSs in SNOW-tree.\label{fig:sim_tree}]{
    \includegraphics[width=0.31\textwidth]{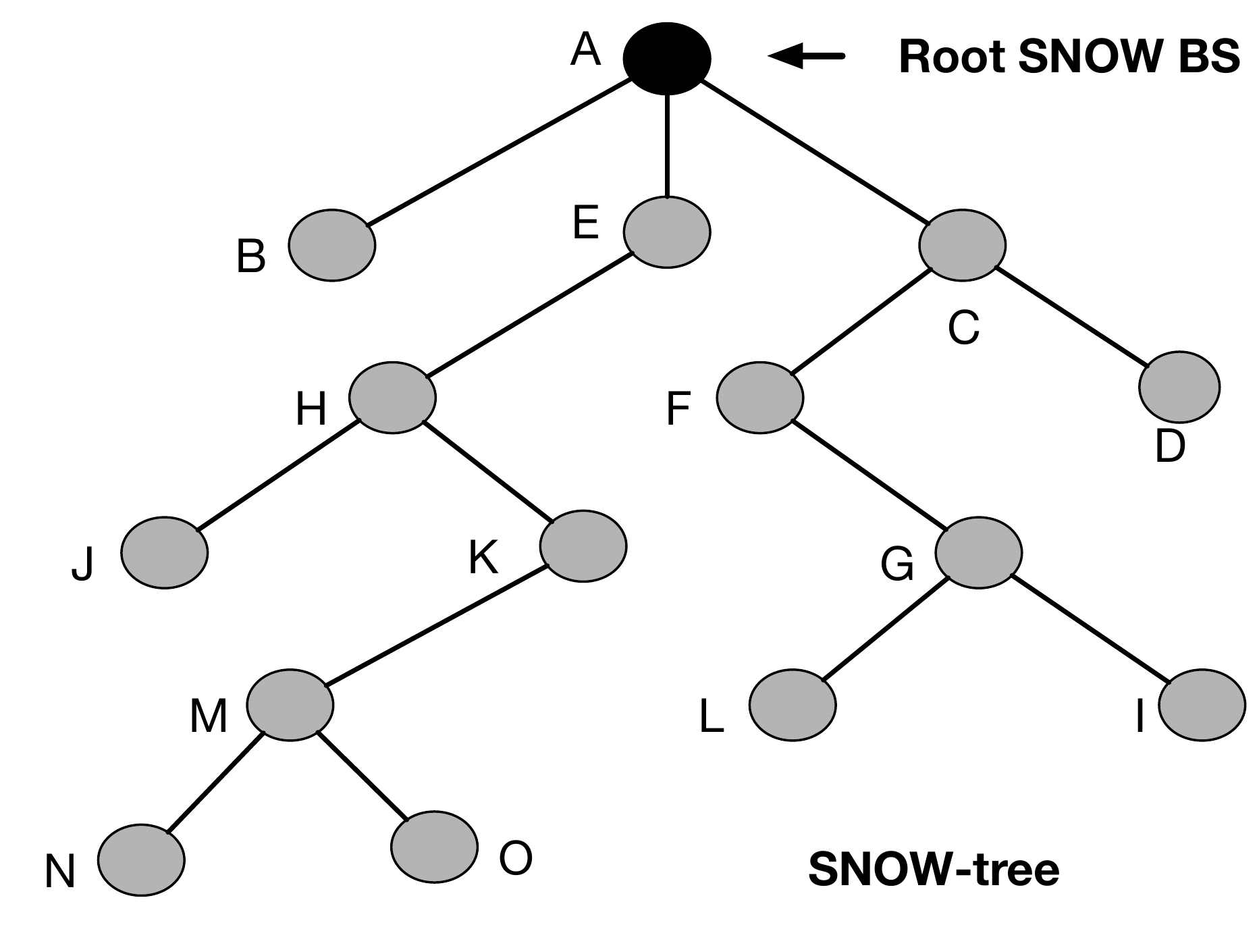}
      }
    \hfill
      \subfigure[Max. allowed number of common subcarriers between interfering SNOW pairs.\label{fig:interf_mat}]{
        \includegraphics[width=.31\textwidth]{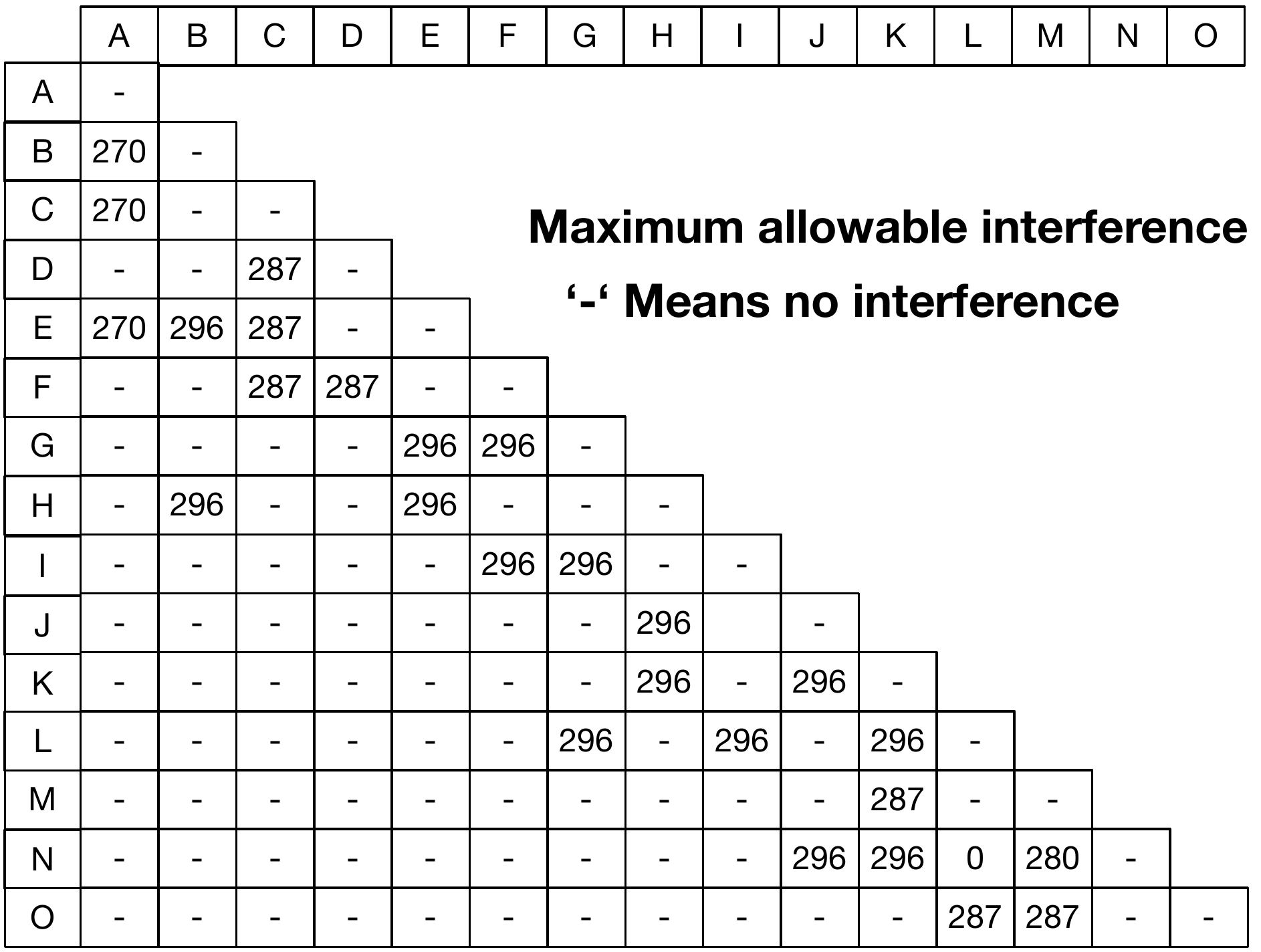}
      }
      \hfill
      \subfigure[Subcarrier assignment by root BS at location A in SNOW-tree.\label{fig:sop_sim}]{
        \includegraphics[width=0.31\textwidth]{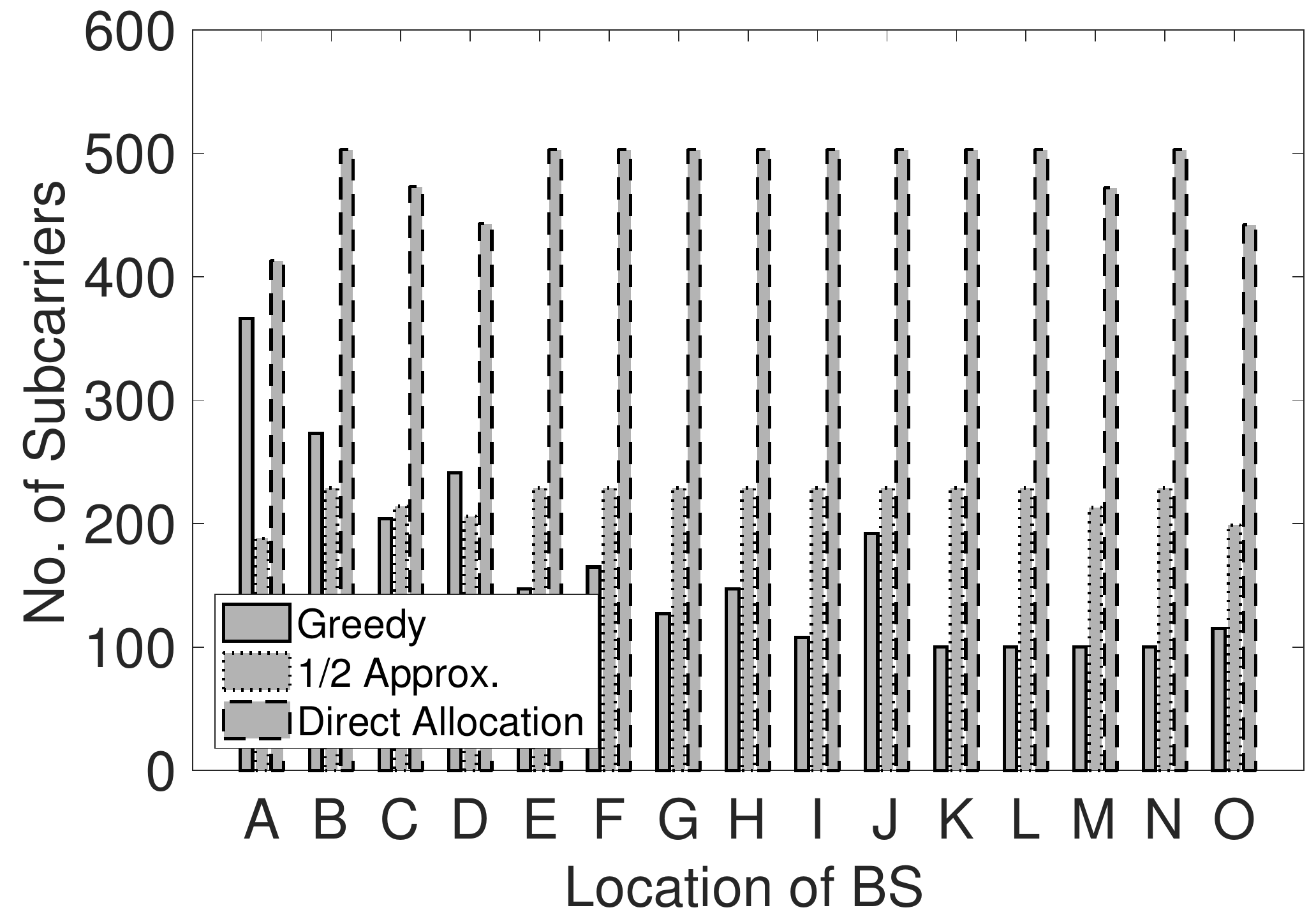}
      }
    \caption{SNOW-tree topology, allowable interference between BSs, and subcarrier allocation for BSs in simulation.}\vspace{-0.15in}
    \label{fig:sim_setup}
\end{figure*}
\begin{figure*}[!htb]
    \centering \vspace{-0.1in}
      \subfigure[Reliability in multi-level inter-SNOW comm.\label{fig:reliability_sim}]{
        \includegraphics[width=.31\textwidth]{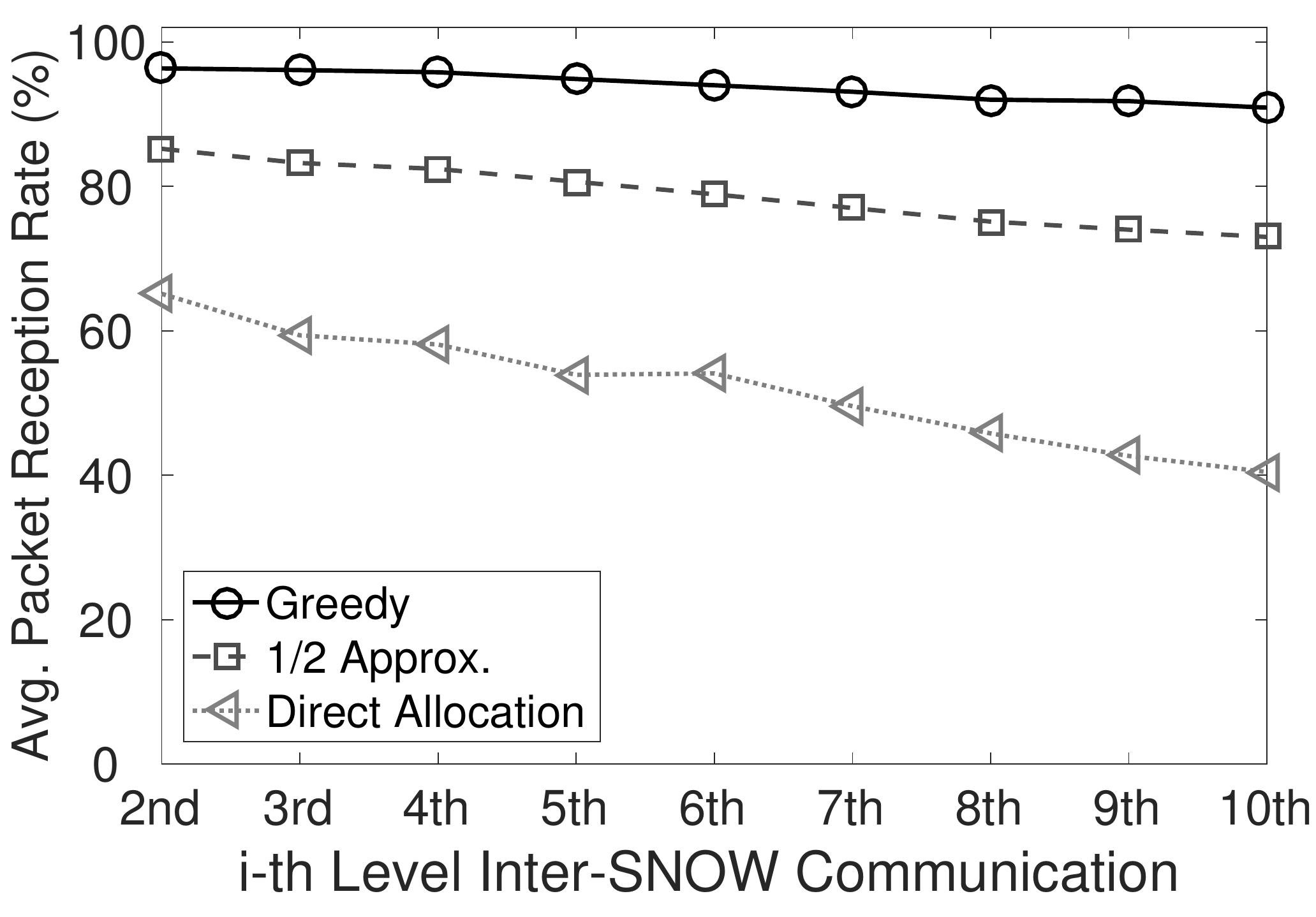}
      }
      \hfill
      \subfigure[Latency in multi-level inter-SNOW comm.\label{fig:latency_sim}]{
        \includegraphics[width=.31\textwidth]{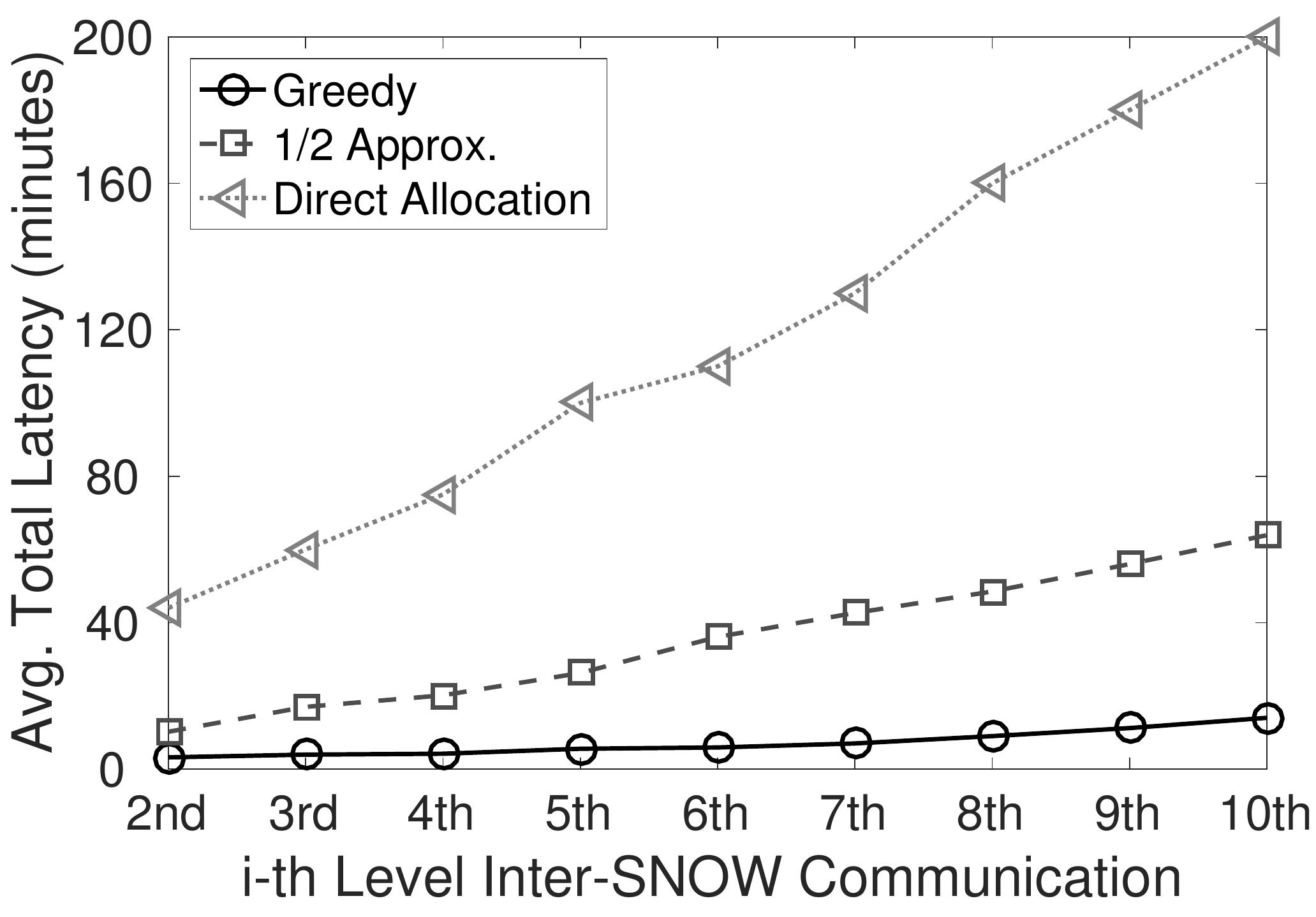}
      }
      \hfill
      \subfigure[Energy in multi-level inter-SNOW comm.\label{fig:energy_sim}]{
        \includegraphics[width=.31\textwidth]{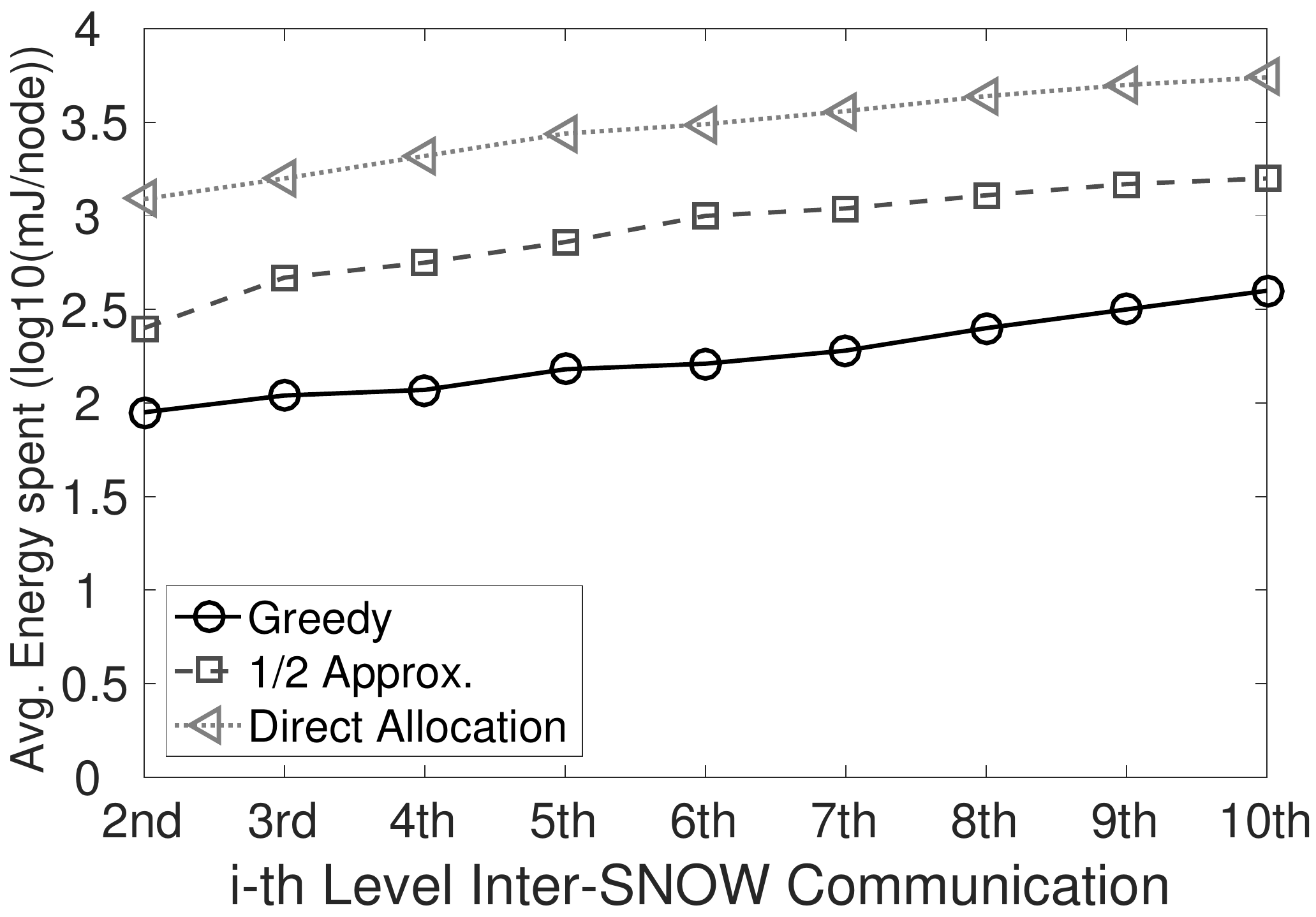}
      }
    \caption{Multi-level parallel inter-SNOW communications in SNOW-tree}\vspace{-0.15in}
    \label{fig:p2p_comm_sim}
\end{figure*}
\subsubsection{Experiments on Inter-SNOW Communications}
To demonstrate inter-SNOW communication performance, we perform parallel communications between two nodes under two sibling BSs in each SNOW-tree, using the sets of subcarriers assigned to BSs by different SOP algorithms in our previous experiments. 
Since, we have only one node under each BS in a tree (as explained in Section~\ref{sec:implementation}), we allow those nodes to use all the subcarriers of their respective BSs.
Considering SNOW-tree 1, the node in BS located at B (and E) will send inter-SNOW packets to the node in BS located at E (and B) via root BS located at A. Thus, this is level three inter-SNOW communication.
In experiments, the node in BS at B (and E) randomly hops into different subcarriers of its BS and sends consecutive 100 packets destined for the node in BS at E (and B). BS at B (and E) first receives the packets (intra-SNOW) and then relays to its parent BS at A (inter-SNOW). Root BS at A then relays (inter-SNOW) the packets to BS at E (and B). Finally, BS at E (and B) sends (intra-SNOW) the packets to its node. Considering a single inter-SNOW packet, since the node is randomly hopping to different subcarriers, the BS sends (intra-SNOW) the same packet via all subcarriers, so that the node may receive it instantly. The whole process is repeated 1000 times in every SNOW-tree.
Figure~\ref{fig:p2p_comm} shows the average PRR, latency, and energy consumption in inter-SNOW communications, while the set of subcarriers used are given by our greedy heuristic, our approximation algorithm, and the direct allocation scheme in our previous experiments in Section~\ref{sec:exp_sop}.

Figure~\ref{fig:p2p_reliability} shows that the average PRR values are high in all SNOW-trees when the subcarriers are assigned using our greedy heuristic. For example, PRR is as high as 99.99\% in SNOW-tree 5 compared to 97.2\% and 74\% by our approximation algorithm and the direct allocation scheme, respectively.
Figure~\ref{fig:p2p_latency} shows that the per inter-SNOW packet latency is lower in all SNOW-trees in case of our greedy subcarrier assignments. In SNOW-tree 5, it is 26.2ms on average compared to 32.8ms and 50ms in cases of our approximation algorithm and the direct allocation scheme assignments, respectively.
Figure~\ref{fig:p2p_energy} shows average energy consumed per inter-SNOW packet at Tx and Rx nodes are lower in all SNOW-trees for our greedy assignments. In SNOW-tree 5, Tx and Rx nodes consume on average 0.49mJ and 0.48mJ energy, respectively. For our approximation algorithm, these values are 0.59mJ and 0.56mJ, while the direct allocation yields 1.2mJ and 1mJ. 
These experiments thus confirm that our greedy heuristic and approximation algorithms are practical choices for SOP.

%% file: simulation.tex
\subsection{Simulation}\label{sec:simulations}
For evaluation under large-scale network, we perform simulations through NS-3~\cite{ns3}.
\subsubsection{Simulation Setup}\label{sec:sim_setup}
We create a SNOW-tree of 15 SNOWs (BSs) as shown in Figure~\ref{fig:sim_tree} and simulate the (25x15)km$^2$ area as shown in Figure~\ref{fig:testbed}. BS at location A is the root BS. Each SNOW has 1000 nodes, totaling 15000 thousand nodes in the SNOW-tree. We limit the maximum allowable number of common subcarriers between interfering BSs based on the white space availability at different BS locations (Figure~\ref{fig:channels}) and our experimental findings, which is shown in Figure~\ref{fig:interf_mat}. $\sigma_i$ in Constraint (\ref{c:const1}) is chosen to be 100 for all the BS. Thus, a subcarrier will be used by at most 10 nodes in worst case in intra-SNOW communication. Figure~\ref{fig:sop_sim} shows the subcarrier assignments for all BSs by the root BS at location A, while using our greedy heuristic algorithm, approximation algorithm, and the direct allocation scheme. Here, both greedy heuristic and approximation algorithms do not violate any of the Constraints of SOP. However, the direct allocation scheme violates Constraints (\ref{c:const2}) and (\ref{c:const3}) of SOP. The values for various parameters such as packet size, spreading factor, modulation, and Tx power are set the same as described in our real experiments (Section~\ref{sec:exp_setup}).

\subsubsection{Simulation Results}

We evaluate the performance of our design using thousands of nodes by generating thousands of parallel multi-level inter-SNOW communications. In simulation, each node in each SNOW sends 100 packets with a random sleep interval of 0-50 ms, destined for another node in second level (adjacent SNOWs) and up to its maximum reachable level inside the SNOW-tree. In each SNOW, we identify nodes from 1 to 1000. In our simulation, a node with ID $i$ will send inter-SNOW packets to the nodes with ID $i$ in all other SNOWs.
Figure~\ref{fig:p2p_comm_sim} demonstrates the performances in terms of reliability, latency, and energy consumption when subcarriers assigned by our greedy heuristic, approximation algorithm, and direct allocation scheme are used.

Figure~\ref{fig:reliability_sim} shows that by using the subcarriers assigned by our greedy heuristic algorithm, we can achieve on average PRR of 93\% even in 10th level inter-SNOW communications. On the other hand, our approximation algorithm and direct allocation scheme can provide approximately 73\% and 40\% of average PRR, respectively.
Figure~\ref{fig:latency_sim} shows that by using the subcarriers assigned by our greedy heuristic algorithm, we observe on average total latency of 14 minutes to send all successful inter-SNOW packets to the second levels and up to the maximum achievable levels by all 15000 nodes. Using subcarriers given by our approximation algorithm and direct allocation scheme, these values are approximately 60 minutes and 200 minutes, respectively.
Figure~\ref{fig:energy_sim} shows that by using the subcarriers assigned by our greedy heuristic algorithm, the per node energy consumption to send all successful inter-SNOW packets to all possible levels is 389mJ. While in cases of our approximation algorithm and direct allocation scheme, these values are 1728mJ and 5580mJ, respectively. Thus, the simulation results demonstrate that the greedy heuristic or the approximation algorithm can be chosen to scale up LPWANs for future IoT applications.

\subsection{Discussion}
In Section~\ref{sec:proposed_solution}, we have justified that our greedy heuristic approach is an intuitive and highly scalable polynomial-time solution. Additionally, we have discussed that deriving an analytical bound (in terms of scalability) of our greedy heuristic is not immediate. Hence, for the cases when an analytical performance bound is needed, we have proposed a probabilistic optimization approach and derived its theoretical performance bound (Section~\ref{sec:bounded_algorithm}). Specifically, our probabilistic optimization approach is a $\frac{1}{2}$-approximation algorithm. 
In terms of performance, both experiments and simulations demonstrate that our greedy heuristic algorithm provides higher reliability, lower latency, and lower energy consumption in both intra- and inter-SNOW communications compared to our approximation algorithm, which is due to its interference-aware subcarrier assignments to different SNOW BSs. Since our approximation algorithm assigns more subcarriers to most of the BSs (both in experiments and simulations), it assigns a greater number of interfering subcarriers between neighboring BSs. Such assignment by our approximation algorithm causes frequent back-offs in transmissions by the nodes, resulting in an increase in latency and energy consumption in both intra- and inter-SNOW communications.

As described in Algorithms~\ref{algo:greedy} and~\ref{algo:approx}, our greedy heuristic or/and approximation algorithms may fail to provide a feasible subcarrier assignment for few SOP problem instances.
In practice, either our greedy heuristic or our approximation algorithm may be adopted to handle the subcarrier assignment failure of each other. In cases when both fail, the target application's requirement will dictate which solution should be adopted. For example, if the application requires bounded performance and high spectrum utilization, our approximation algorithm may be adopted. On the other hand, greedy heuristic may be chosen in case higher reliability is expected. In experiments, we were unable to demonstrate such cases based on the available TV white spaces and environments at our testbed location. Our realistic simulations, where parameters are chosen based on our experiments, do not also showcase any infeasible cases of our greedy heuristic algorithm. In general, our experiments and simulations demonstrate that both greedy heuristic and approximation algorithms may be practically chosen to scale up LPWANs for future IoT applications.